\documentclass[a4paper]{grant_nature}

\usepackage{graphicx}
\graphicspath{{./}{figures/}}
\usepackage{amsmath,amssymb}
\usepackage{xspace}
\usepackage{color,hyperref}
\definecolor{linkcolor}{rgb}{0,0.3,0.7}
\hypersetup{colorlinks=true,
	linkcolor=linkcolor,
	citecolor=linkcolor,
	filecolor=linkcolor,
	urlcolor=linkcolor}
\usepackage{lineno}
\usepackage{footmisc} 
\newcommand{\gfo}{AT2017gfo\xspace}

\newcommand\blfootnote[1]{%
  \begingroup
  \renewcommand\thefootnote{}\footnote{#1}%
  \addtocounter{footnote}{-1}%
  \endgroup
}

\title{Identification of strontium in the merger of two neutron stars}

\author{Darach Watson$^{\ref{af:NBI},\ref{af:DAWN}}$,
Camilla~J. Hansen$^{\ref{af:MPIA_Heidelberg},*}$,
Jonatan Selsing$^{\ref{af:NBI},\ref{af:DAWN},*}$,
Andreas Koch$^{\ref{af:U_Heidelberg}}$,
Daniele B. Malesani$^{\ref{af:NBI},\ref{af:DAWN},\ref{af:DTU_space}}$,
Anja~C. Andersen$^{\ref{af:NBI}}$,
Johan~P.~U. Fynbo$^{\ref{af:NBI},\ref{af:DAWN}}$,
Almudena Arcones$^{\ref{af:TU_Darmstadt},\ref{af:GSI_Darmstadt}}$,
Andreas Bauswein$^{\ref{af:GSI_Darmstadt},\ref{af:HITS}}$,
Stefano Covino$^{\ref{af:Brera}}$,
Aniello Grado$^{\ref{af:INAF-OACN}}$,
Kasper~E. Heintz$^{\ref{af:NBI},\ref{af:DAWN},\ref{af:U_Iceland}}$,
Leslie Hunt$^{\ref{af:INAF-OAD}}$,
Chryssa Kouveliotou$^{\ref{af:GWU},\ref{af:APSIS}}$
Giorgos Leloudas$^{\ref{af:NBI},\ref{af:DTU_space}}$,
Andrew Levan$^{\ref{af:Radboud},\ref{af:U_Warwick}}$,
Paolo Mazzali$^{\ref{af:LJMU},\ref{af:MPIA_Garching}}$,
Elena Pian$^{\ref{af:INAF-ASSOB}}$
{\footnotesize[See end for affiliations]}
}

\begin{document}
\maketitle

\begin{abstract}
  Half of all the elements in the universe heavier than iron were created by
  rapid neutron capture. The theory for this astrophysical `\emph{r}-process'
  was worked out six decades ago and requires an enormous neutron flux to make
  the bulk of these elements.\cite{BBFH1957} Where this happens is still
  debated.\cite{2019Natur.569..241S} A key piece of missing evidence is the
  identification of freshly-synthesised \emph{r}-process elements in an
  astrophysical site. Current
  models\cite{Lattimer1977,Eichler1989,Freiburghaus1999} and circumstantial
  evidence\cite{Ji2016} point to neutron star mergers as a probable
  \emph{r}-process site, with the optical/infrared `kilonova' emerging in the
  days after the merger a likely place to detect the spectral signatures of
  newly-created neutron-capture
  elements.\cite{Metzger2010,Barnes&Kasen2013,Tanvir2013} The kilonova, \gfo,
  emerging from the gravitational-wave--discovered neutron star merger,
  GW170817,\cite{2017PhRvL.119p1101A} was the first kilonova where detailed
  spectra were recorded. When these spectra were first
  reported\cite{2017Natur.551...67P,2017Natur.551...75S} it was argued that they
  were broadly consonant with an outflow of radioactive heavy elements, however,
  there was no robust identification of any element. Here we report the
  identification of the neutron-capture element strontium in a re-analysis of
  these spectra. The detection of a neutron-capture element associated with the
  collision of two extreme-density stars establishes the origin of
  \emph{r}-process elements in neutron star mergers, and demonstrates that
  neutron stars comprise neutron-rich matter\cite{Baade&Zwicky1934}.\end{abstract}

The most detailed information available for a kilonova comes from a series of
spectra of \gfo taken over several weeks with the medium resolution, ultraviolet
(320\,nm) to near-infrared (2,480\,nm) spectrograph, X-shooter, mounted at the
Very Large Telescope at the European Southern Observatory. These
spectra\cite{2017Natur.551...67P,2017Natur.551...75S},
allow us to track the evolution of the kilonova's primary electromagnetic output
from 1.5 days until 10 days after the event. Detailed modelling of these spectra
has yet to be done owing to the limited understanding of the phenomenon and the
expectation that a very large number of moderate to weak lanthanide lines with
unknown oscillator strengths would dominate the
spectra\cite{Tanaka2013,2017Natur.551...80K}. Despite the expected complexity,
we sought to identify individual elements in the early spectra because these
spectra are well-reproduced by relatively simple
models\cite{2017Natur.551...67P}.

\begin{figure}[!ht]
 \includegraphics[width=\columnwidth,clip=]{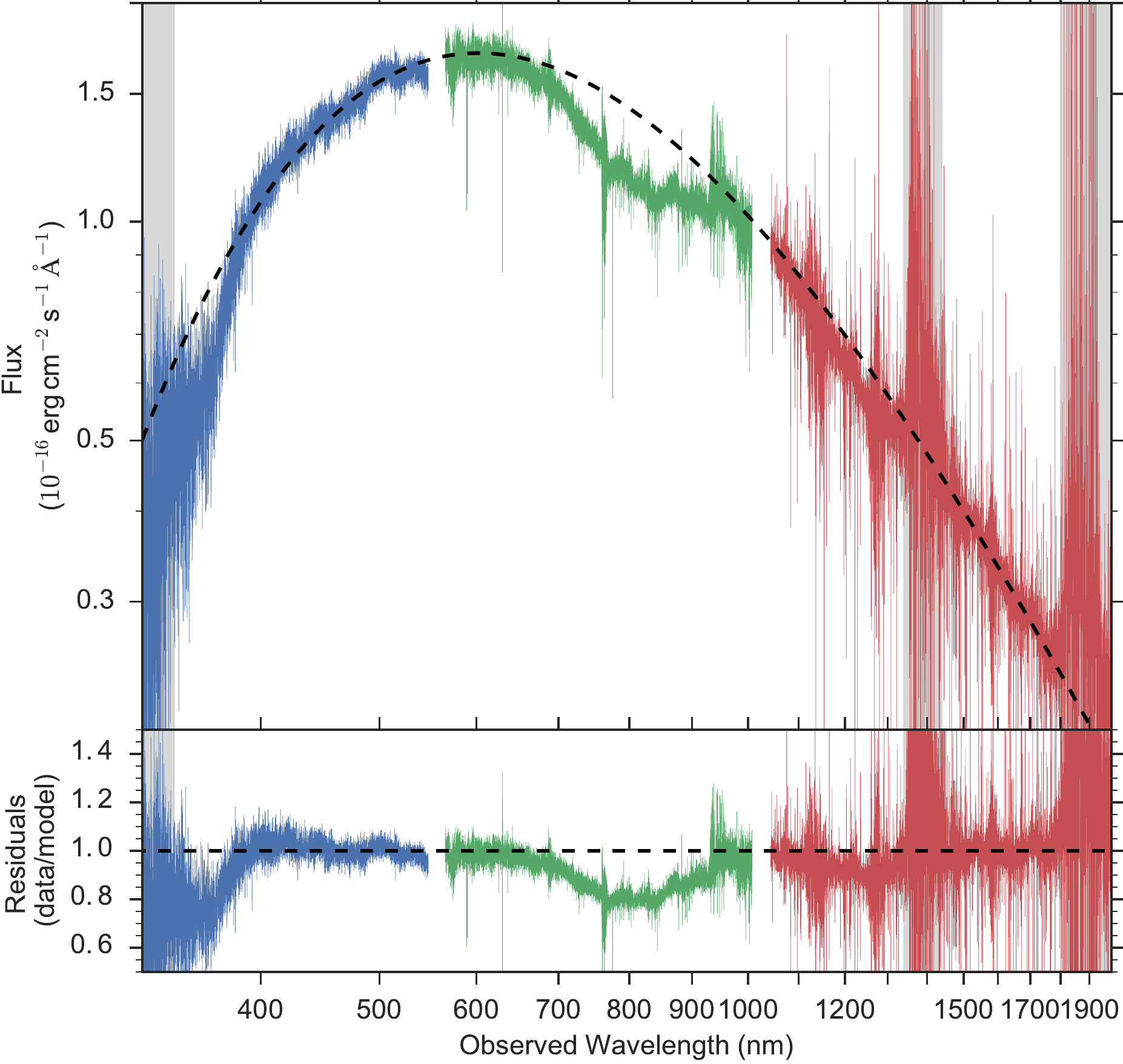}
 \caption[]{\textbf{Spectrum of the kilonova, \gfo,
            showing broad absorption features.} The spectrum was observed at
            1.5 days after the merger. The
            dashed line is the blackbody component of a
            blackbody model with broad absorption lines (see text).
            The residuals of data minus blackbody are shown in
            the lower panel with the dashed line indicating the $1\sigma$
            uncertainty on each spectral bin. The data in the sections
            overplotted with grey bars are
            affected by telluric features or are poorly-calibrated regions
            and are not included in the fit.
            \label{fig:first_epoch_spectrum}
           }
\end{figure}

The first epoch spectrum can be reproduced over the entire observed spectral
range with a single-temperature blackbody with an observed temperature
\(\simeq 4,800\)\,K. The two major deviations
short of $1\,\mu$m from a pure blackbody are due to two very broad ($\sim0.2c$)
absorption components. These components are observed centred at about 350\,nm
and 810\,nm (Fig.~\ref{fig:first_epoch_spectrum}). The shape of the ultraviolet
absorption component is not well constrained because it lies close to the edge
of our sensitivity limit and may simply be cut off below about 350\,nm. The
presence of the absorption feature at 810\,nm at this epoch has been noted in
earlier publications\cite{2017Natur.551...75S,2017Natur.551...67P}.

The fact that the spectrum is very well reproduced by a single temperature
blackbody in the first epoch suggests a population of states 
close to local
thermal equilibrium (LTE). We therefore used three separate methods of
increasing complexity initially to determine, without too many assumptions, the
most likely origin of the spectral features, and then to self-consistently model
and test our conclusion. First, our own LTE spectral synthesis code, second, the
LTE line analysis and spectrum synthesis code
MOOG\cite{2012ascl.soft02009S}, and third the moving plasma
radiative transfer code, TARDIS\cite{Kerzendorf2014} (see Methods). We used a
variety of spectral line lists for these codes, all of which yielded consistent
results. For our own LTE code, we adopted a fiducial temperature of 3,700\,K,
which is our final model's best-fit temperature corrected by the Doppler factor
($-0.23$) of the absorption features we determine below; changing the
temperature of our LTE model in the range 3,700--5,100\,K does not significantly
affect our results.

To identify the absorption features, we seek lines with wavelengths
blueshifted by 0.1--0.3$c$, corresponding approximately to
390--500\,nm and 900--1,160\,nm in the rest frame (see Methods).
The lines will also be broadened with an observed width dependent
on the velocity and geometry. For spherically expanding ejecta, the
line-broadening will be similar to the expansion velocity of
the gas. We do not attempt a detailed geometric model here because
it depends on assumptions about the geometry of the gas and the
wavelength-dependent opacity, with significant relativistic and
time-delay corrections.

\begin{figure}
 \includegraphics[width=\columnwidth,clip=]{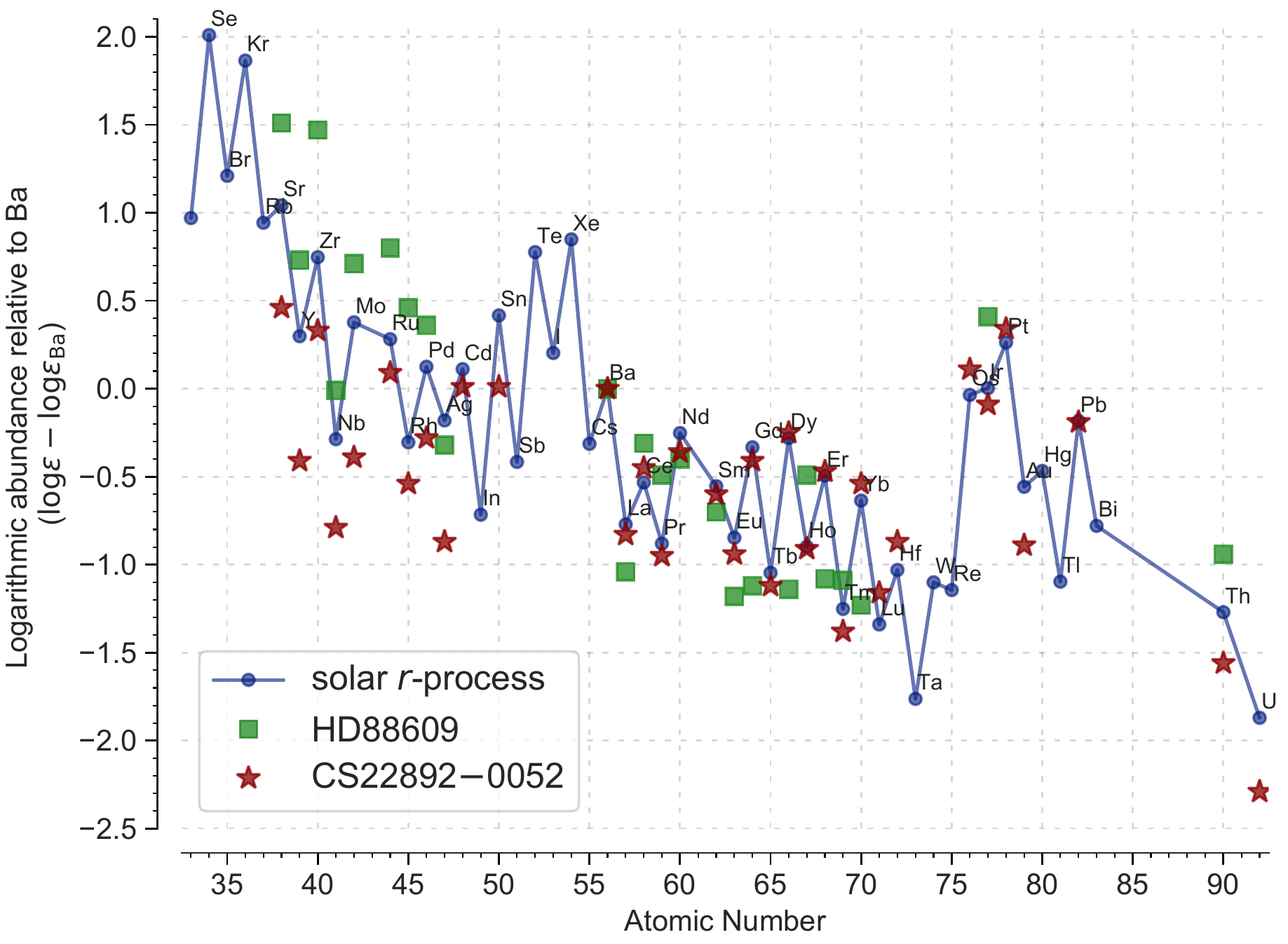}
 \caption[]{\textbf{The abundances of elements produced by the \emph{r}-process.}
            Relative \emph{r}-process abundances normalized to the Ba
            abundance are shown for the sun and two metal-poor stars, one
            rich in heavy \emph{r}-process elements,
            CS\,22892$-$052\cite{Sneden2003,Sneden2009}, and the other rich
            in light \emph{r}-process elements, HD\,88609\cite{Honda2007}.
            These are the abundances of the elements used in
            Fig.~\ref{fig:model_spectrum}, \emph{inset}.
            \label{fig:r-process_abundances}
           }
\end{figure}

We adopt an initially agnostic view on the expected abundances. We use solar
\emph{r}-process abundance ratios (total solar abundances of heavy
elements\cite{Lodders2011} with \emph{s}-process elements
subtracted\cite{2014ApJ...787...10B}), and abundances from two metal-poor
stars, old enough to be dominated by the \emph{r}-process in their
neutron-capture abundances\cite{Honda2007,2000ApJ...533L.139S}.  These three
sets span a wide range in the ratio of light to heavy \emph{r}-process
abundances (Fig.~\ref{fig:r-process_abundances}).  We also produce
absorption spectra for each element individually (Extended Data
Figs.~\ref{exfig:moog_individual_spectra_plot} and
\ref{exfig:LTE_individual_spectra}).

\begin{figure}
 \includegraphics[width=\columnwidth,clip=]{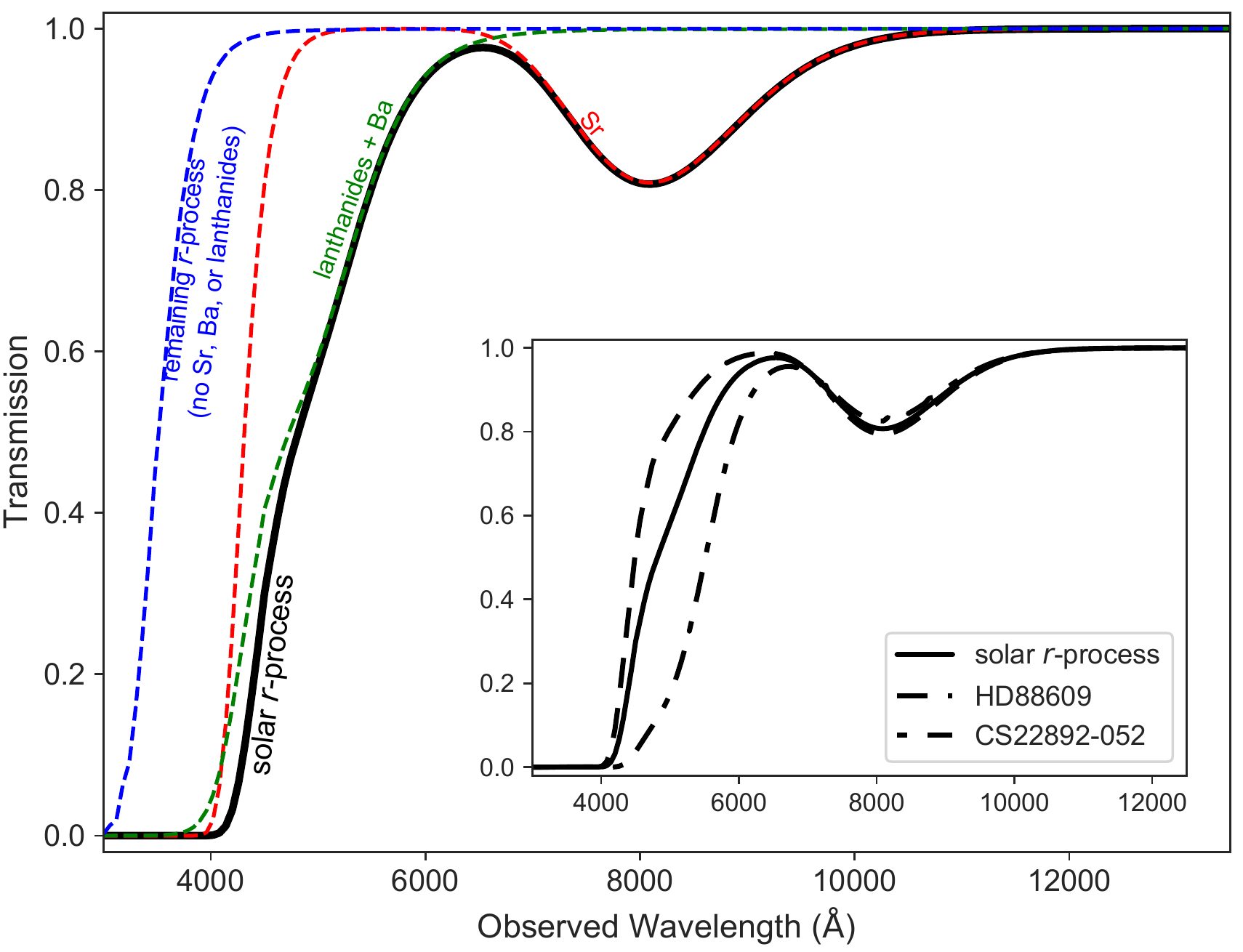}
 \caption[]{\textbf{Thermal \emph{r}-process element transmission spectrum.} The spectra are
            based on the lines formed in a gas in local thermal equilibrium
            with a temperature of 3,700\,K and an electron
            density of $10^7$\,cm$^{-3}$, broadened by $0.2c$ and
            blueshifted by $0.23c$. The spectrum produced by a solar
            \emph{r}-process abundance ratio is plotted as a solid line.
            Contributions due to Sr (red dashed),
            Ba and the lanthanides (green dashed), and the
            remaining \emph{r}-process elements (blue dashed) are shown.
            \emph{Inset:} spectra resulting from a solar \emph{r}-process abundance
            ratio (solid line), and from the abundance ratios of the
            metal-poor stars HD\,88609\cite{Honda2007} (dashed line) and
            CS\,22892$-$052\cite{Sneden2003,Sneden2009} (dash-dotted line).
            \label{fig:model_spectrum}
           }
\end{figure}

Our LTE models with abundances from a solar-scaled \emph{r}-process and
metal-poor star abundances all show that Sr produces a strong feature  centred
at an observed wavelength of $\sim800$\,nm, as well as features shortward of
$\sim400$\,nm, for our adopted blueshift (Fig.~\ref{fig:model_spectrum}, see
also Extended Data Fig.~\ref{exfig:LTE_individual_spectra}).  The restframe
wavelengths of the longer wavelength features are 1,000--1,100\,nm. It is worth
noting that Sr is typically considered an \emph{s}-process element because only
about 30\% of the cosmic (solar) abundance is produced by the
\emph{r}-process\cite{Lodders2011,2014ApJ...787...10B}. For this reason it has
not always been considered in kilonova simulations. However, it is one of the
more abundant \emph{r}-process elements, accounting for at least a few percent
by mass of all \emph{r}-process elements\cite{2014ApJ...787...10B}. Of all the
\emph{r}-process elements Sr displays by far the strongest absorption features
in this region of the spectrum (Extended Data
Figs.~\ref{exfig:moog_individual_spectra_plot} and
\ref{exfig:LTE_individual_spectra}). The lanthanide elements, and especially Ba,
produce strong absorption but only in the optical region shortward of about
650\,nm.  The spectral features we observe can therefore only be Sr, an element
produced near the first \emph{r}-process peak.

The 810\,nm feature was previously
proposed\cite{2017Natur.551...75S} to  originate in absorption from
Cs\,\textsc{i} and Te\,\textsc{i}. This identification can be
ruled out because neither Cs\,\textsc{i} nor Te\,\textsc{i} produce strong lines in a plasma at this temperature (Extended Data
Fig.~\ref{exfig:LTE_individual_spectra}). Much stronger lines would be  expected from ions
of other elements co-produced with Cs (Z=55, e.g.\ from
La\,\textsc{ii}, see Methods).

The most abundant \emph{r}-process elements are those in the first peak
(Fig.~\ref{fig:r-process_abundances}), elements
with $A\sim80$, and of these, it is Sr, Y (Z=39), and Zr (Z=40) that are
easily detected in a low density, $\sim4,000$\,K thermal plasma, because
these elements have low excitation potentials for their singly-charged ions.
Seen in this context, the detection of Sr in \gfo is not surprising, in spite of prior
expectations that the spectra would be dominated by heavier
elements\cite{Kasen2013,Tanaka2013}. Furthermore, the atomic levels in Sr
that give the absorption lines observed at $810$\,nm are metastable.
Photo-excitation can increase the population in these states, strengthening
the $810$\,nm feature significantly\cite{1990sjws.conf..149J} compared to
resonance blue/near-ultraviolet absorption lines. Ba and the lanthanide series contribute
significantly to the total opacity of \emph{r}-process material in the
optical region of the spectrum (Fig.~\ref{fig:model_spectrum}). But we do not
detect strong optical features.  We cannot, however, easily exclude the
presence of $A\gtrsim140$ elements on this basis. Even if we
could exclude the presence of heavier elements in the outer layers of the
thermal, expanding cloud, there is no way from these early spectra to exclude
that such elements could exist at lower depths or in an obscured
component.

\begin{figure}
 \centering
 \includegraphics[width=\columnwidth,clip=]{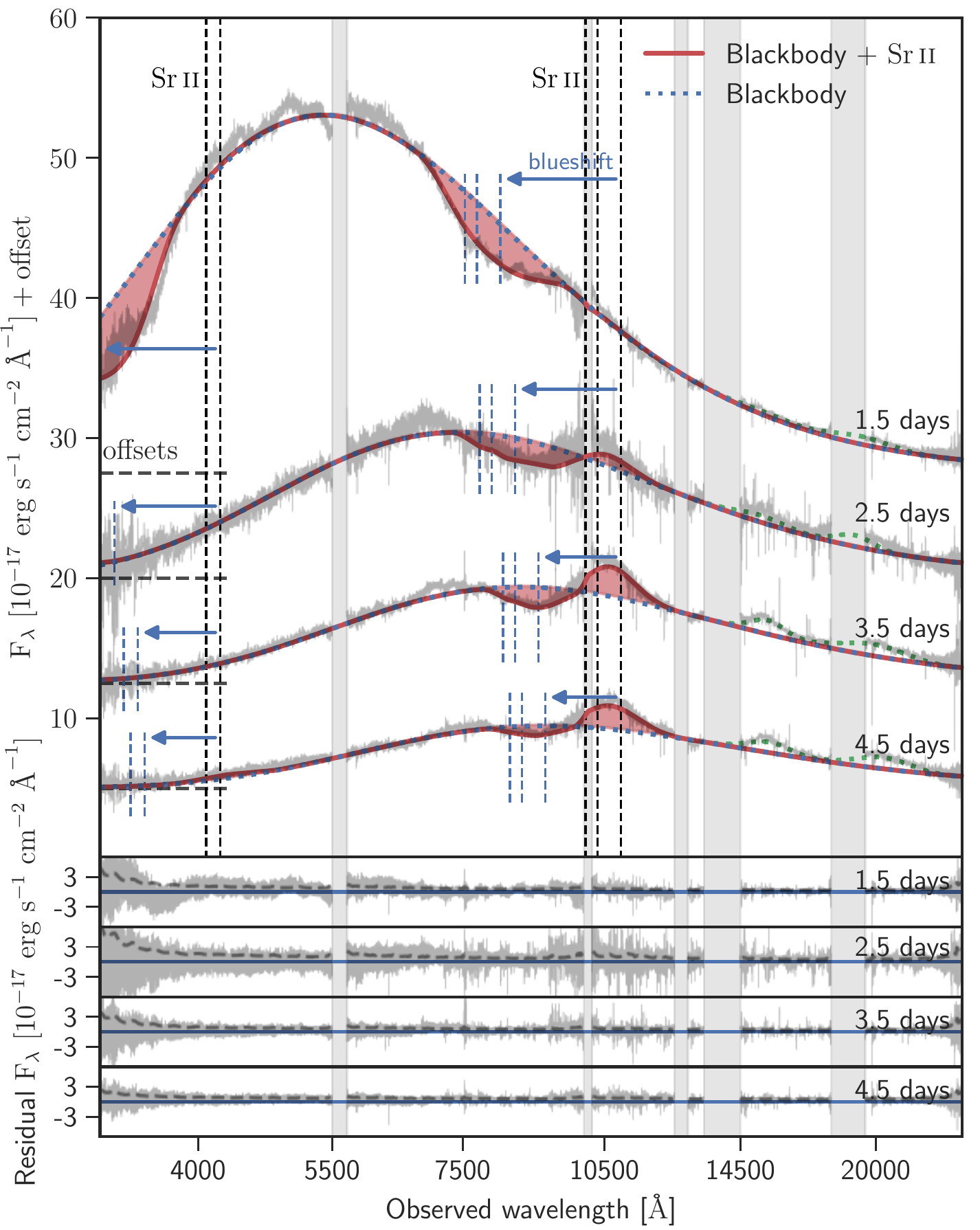}
 \caption{\textbf{Spectral series of \gfo 1.5--4.5\,days after the
          merger.} Data are shown in
          grey and have been smoothed slightly.
          A model (solid red lines) consisting of a blackbody
          (blue dotted lines) with P\,Cygni profiles
          (red transparent fill) for the Sr lines is shown.
          The rest (black) and observed (blue) positions of the
          model's Sr lines are shown, with the
          blueshift indicated by arrows.
          Green dotted lines show the Gaussian emission profiles
          added to ensure the overall continuum
          is not biased. A vertical offset has been applied to
          each spectrum
          for clarity, with zero flux indicated
          by the dashed horizontal line segment. Bottom panels
          show the residuals between model and data.
          \label{fig:pcygni_stack}
         }
\end{figure}

Since a simple \emph{r}-process abundance LTE model can account
well for the  first epoch spectrum, we expand it to the subsequent
three epochs while the kilonova is still at least partially
blackbody-like. With a freely expanding explosion we expect to
begin observing P\,Cygni lines once the  outer absorbing
`atmosphere' begins to become more optically thin and attain a
significant physical radius with respect to the photospheric
radius. We fit the first four epochs as a blackbody with P\,Cygni
lines from Sr. We fit only the strongest lines to reduce our
computational time to a manageable level, as these lines
provide most of the  opacity at these wavelengths. These fits are
shown in  Fig.~\ref{fig:pcygni_stack} and offer a compelling
reproduction of the spectra at all three epochs. The P\,Cygni
model has free parameters for the velocities of the photosphere
and atmosphere, which change the shape of the profile. The fit is
remarkable given its simplicity and our lack of knowledge of the
system geometry. We note that P\,Cygni  emission components are
always centred close to the rest wavelength of the spectral lines,
so the observed wavelength of the emission line is not a free
parameter. The most prominent emission component observed
throughout the spectral series is centred close to 1,050\,nm, and
the weighted restframe centre of the near-infrared lines from Sr
is also 1,050\,nm. This adds to our  confidence in the line
identification based on the simple thermal \emph{r}-process
absorption model.

We further confirm our results using TARDIS, extending the code's atomic database to
include elements up to $_{92}$U with the latest
Kurucz linelists\cite{Kurucz2017} with its 2.31 million lines.  Our TARDIS models produce
results very similar to our static-code models, reproducing the spectra well
(Extended Data Fig.~\ref{exfig:tardis_model}).  In particular, the P\,Cygni emission/absorption
structure is well-reproduced as expected, confirming our LTE and MOOG
modelling, and showing Sr dominating the features around $1\,\mu$m.

From the detection of Sr, it is clearly important to consider lighter \emph{r}-process
elements in addition to the lanthanide
elements in shaping the kilonova emission spectrum.
Observations of abundances in stars in dwarf
galaxies\cite{Ji2016} suggest that large amounts of Sr are produced together
with Ba (Z=56) in infrequent events, implying the existence of a site that
produces both light and heavy \emph{r}-process elements together in quantity, as
found in some models\cite{Wanajo2014,Just2015}. This is consistent
with our spectral analysis of \gfo and analyses of its
lightcurve\cite{Drout2017,2017ApJ...848L..27T}. Together with the
differences observed in the relative abundances of \emph{r}-process Ba and Sr
in stellar spectra\cite{2014ApJ...797..123H}, this suggests that the relative
efficiencies of light and heavy \emph{r}-process production could vary
substantially from merger to merger.

Extreme-density stars composed of neutrons were proposed shortly after the
discovery of the neutron\cite{Baade&Zwicky1934}, and identified with pulsars
three decades later\cite{Hewish1968}. However, no spectroscopic confirmation of
the composition of neutron stars has ever been made. The identification here of
an element that could only have been synthesised so quickly under an extreme
neutron flux, provides the first direct spectroscopic evidence that neutron
stars comprise neutron-rich matter.

\blfootnote{%
\begin{affiliations}
\descitem{af:NBI}Niels Bohr Institute, University of Copenhagen, Copenhagen, Denmark.
\descitem{af:DAWN}Cosmic Dawn Center (DAWN).
\descitem{af:MPIA_Heidelberg}Max-Planck-Institut f\"ur Astronomie, Heidelberg, Germany.
\descitem{af:U_Heidelberg}Zentrum f\"ur Astronomie der Universit\"at Heidelberg, Astronomisches Rechen-Institut, Heidelberg, Germany.
\descitem{af:DTU_space}DTU Space, National Space Institute, Technical University of Denmark, Kgs.\ Lyngby, Denmark.
\descitem{af:TU_Darmstadt}Institut f\"ur Kernphysik, Technische Universit\"at Darmstadt, Darmstadt, Germany.
\descitem{af:GSI_Darmstadt}GSI Helmholtzzentrum f\"ur Schwerionenforschung GmbH, Darmstadt, Germany.
\descitem{af:HITS}Heidelberger Institut f\"ur Theoretische Studien, Heidelberg, Germany.
\descitem{af:Brera}INAF / Brera Astronomical Observatory, Merate, Italy.
\descitem{af:INAF-OACN}INAF-OACN, Naples, Italy.
\descitem{af:U_Iceland}Centre for Astrophysics and Cosmology, Science Institute, University of Iceland, Reykjav\'{\i}k, Iceland.
\descitem{af:INAF-OAD}INAF - Osservatorio Astrofisico di Arcetri, Firenze, Italy.
\descitem{af:GWU}Physics Department, the George Washington University, Physics Department, Washington, DC, USA.
\descitem{af:APSIS}Astronomy, Physics and Statistics Institute of Sciences (APSIS), Washington, DC, USA.
\descitem{af:Radboud}Department of Astrophysics/IMAPP, Radboud University Nijmegen, The Netherlands.
\descitem{af:U_Warwick}Department of Physics, University of Warwick, Coventry, UK.
\descitem{af:LJMU}Astrophysics Research Institute, Liverpool John Moores University, Liverpool, UK.
\descitem{af:MPIA_Garching}Max-Planck Institute for Astrophysics, Garching, Germany.
\descitem{af:INAF-ASSOB}INAF, Astrophysics and Space Science Observatory, Bologna, Italy.
$^*$These authors contributed equally to this work.
\end{affiliations}
}%

\clearpage

\begin{methods}

\subsection{Spectral synthesis}\label{sec:spectral_synthesis}

Different codes were used to compute synthetic absorption spectra. We used
MOOG\cite{2012ascl.soft02009S,MOOG_software} v.~2014, and our own single temperature and
density LTE code. In addition, we verified our results using the TARDIS
supernova spectral synthesis code. For the first two codes we used line
lists gathered from the literature. These line lists can be seen in 
Supplementary Data~1. For the TARDIS modelling we used the line lists of
Kurucz. Our codes yield consistent results with the different line lists.

MOOG is a synthetic spectrum code normally used to
generate synthetic absorption spectra of photospheres in cool stars under the
assumption of local thermodynamic equilibrium. It requires a model atmosphere
that dictates how temperature, gas pressure, and electron density behave in
different layers of the surface gas. Here we adopt Kurucz model
atmospheres\cite{2004astro.ph..5087C}. The second requirement is a line list
containing the rest wavelength of the absorption transition, the element or ion
in which the transition takes place, the excitation potential of the lower
level, and the oscillator strength. The atomic data are based on
refs.~\cite{Biemont2003lines,DenHartog2003Nd,Lawler2001La,Lawler2001Eu,Lawler2001Tb,Lawler2006Sm,Sneden2009}
with updates from NIST.
The strength of the absorption features are calculated
solving radiative transfer equations with a plane parallel treatment of the
atmospheres, assuming that the velocity distribution is Maxwellian, and that
excitations and ionisations are described by the Boltzmann and Saha equations,
respectively. The line/wing damping follows a scaled Uns\"old approximation and
the source function follows a simple black body while scattering (on H, He, and
$e^-$) enters mainly through opacity terms.

Our own code assumes only a gas in LTE without scattering, and that the Boltzmann and Saha
equations can be used to get the ionisation and excitation state of each
element individually. We then use the line lists above and level
information from NIST to determine the relative strengths of the lines.
We adopted a fiducial electron density of $\log{n_e} = 7.8$, based on the
mean density of $0.04\,M_\odot$ of singly-ionized material in a sphere with
the area of the best-fit blackbody. The density of the atmosphere is almost
certainly lower than this.

To make sure the MOOG models and our LTE calculations are reasonably
comparable, for the MOOG models an
effective temperature at the surface of the photosphere of
T$_{\rm eff}\sim5,500$\,K and a surface gravity of \(\log{g}=0\) following
temperature and density profiles in Kurucz model atmospheres, give rise to a
temperature of 3,800\,K and an electron density of $n_e=10^7$\,cm$^{-3}$
within the photosphere.
Absorption lines from lanthanide ions
are believed to be an important source of opacity
due to transitions with unknown oscillator strengths.
For an LTE plasma,
it is likely that such lines are important
and create a complex continuum\cite{Kasen2013,2017Natur.551...80K}.
However, the lanthanide opacity is extremely high in the ultraviolet and blue regions of the spectrum.
The fact that we detect blue emission in the spectrum of \gfo
is already a strong indication
that lanthanide elements do not dominate the early continuum spectrum,
as suggested by other authors\cite{McCully2017,Chornock2017}.
Furthermore, the infrared feature arises from levels that may be overpopulated
due to optical pumping, enhancing the strength of this feature further with
respect to the line-generated continuum at these wavelengths.

Synthetic spectra are generated using both codes based on line lists
containing \emph{r}-process elements, and which are capable of producing strong
features in an LTE plasma at these temperatures. We include all elements
from $_{33}$As up to $_{83}$Bi as well as $_{90}$Th and $_{92}$U. We do not
include the elements $_{35}$Se, $_{36}$Br, $_{37}$Kr, $_{53}$I, or
$_{54}$Xe as they produce no strong or moderate lines at these temperatures
and are rarely detected in stellar spectra\cite{2010IAUS..265...46S}. These
elements have first excitation energies above 5.97\,eV for their neutral
and singly charged ions, giving a fractional population $<10^{-8}$ at our
fiducial temperature. Neither do we include elements with no stable
isotopes ($_{43}$Tc and $_{61}$Pm), nor any molecules. The absorption
line profiles are dominated by the velocity and density distribution of the
expanding atmosphere.

Our line lists contain the strongest lines for LTE spectra at these
temperatures. Since we are interested in finding strong, isolated lines,
this procedure should effectively capture all lines that could
realistically be candidates.

\subsection{Could large numbers of weak lines dominate the optical/NIR opacity?}

The opacity of the kilonova is dominated by absorption lines.
The list of lines we use for MOOG (see references above) has most of the strong
lines in common with the Kurucz list\cite{Kurucz_gfall} we use for the TARDIS modelling. The
results we retrieve from the different techniques and line lists are a useful
check on the robustness of the modelling methodologies. Both methods yield
consistent results, indicating that the overall result presented here is robust
to the selection of the specific line list and the modelling method chosen.
We note that a feature at about 810\,nm is also produced in the spectral
synthesis analysis of ref.~\cite{Tanaka2013}, where lists comprising known lines
are also used. This feature (M.~Tanaka, private communication) is produced
primarily by the same Sr\,\textsc{ii} lines we identify in this work.

The major caveat in identifying line features is the possibility that missing
lines could significantly influence the broad spectral shape compared to
what is predicted from known
lines. Of particular concern are the large numbers of unknown lines from the
lanthanide elements that are likely to dominate the line-expansion opacity\cite{Kasen2013,Tanaka2018}.
While we argue here that our line lists are reasonably complete in strong lines
at these temperatures and densities (and since they are used for modelling stars
with similar temperatures and densities, this makes sense), it is possible that
a very large number of weaker lines could contribute.

However, the line-forming region of the kilonova is likely to be physically
extended, covering a significant fraction of the kilonova radius, particularly
in the near-infrared. The presence of a P\,Cygni profile at $\sim1\,\mu$m
supports that a substantial volume (though not mass) of the kilonova must be
substantially optically thin at this wavelength. The mass absorption
coefficient of the Sr\,\textsc{ii} lines at $\sim1.05\,\mu$m peaks at about
$4\times10^3\,$cm$^2\,$g$^{-1}$ for lines with FWHM$ =0.01\,c$, a temperature of
$5,000\,$K and a density of $10^{-13}\,\mathrm{g}\,\mathrm{cm}^{-3}$. This is at
least two orders of magnitude higher than the mean value for lanthanides such as
Ce and Nd in the optically-thin limit using the Kurucz line lists. Given that
the line lists for these elements are likely to be very incomplete at these
wavelengths, we extrapolate the value of the Ce line opacity of the VALD lines
at 9,000\,\AA\ to $\sim1.05\,\mu$m, which should give a similar opacity
to the line lists calculated in ref.~\cite{Kasen2013} with the
\texttt{autostructure} code. When the lines are extremely optically thick,
within the bulk of the kilonova in the first days, the
Ce opacity is about $10\,$cm$^2\,$g$^{-1}$ (cf. ref.~\cite{Kasen2013}). In the
optically-thin regime in the
outer layers, the Ce line opacity rises about two orders of magnitude.
Using this optically-thin extrapolation of the Ce
lines, the Sr\,\textsc{ii} opacity is still a factor of 4--5 higher, not
including abundance effects that are likely to make the Sr line stronger still.
We show an example of this effect by calculating the expansion opacity for a low
optical depth plasma in Extended Data Fig.~\ref{exfig:expansion_opacity_plot}.
That calculation is purely illustrative, showing how the Sr lines can dominate
the opacity when the gas has low optical depth. For a self-consistent model
calculation, see the TARDIS model spectra in Extended Data Fig.~\ref{exfig:tardis_model}.

\subsection{Spectral modelling}
What appears as two separate emission components are identified in the
spectra. First, a nearly blackbody spectrum modified by absorption
features, that appears to cool over time. Second, an emission component at
redder wavelengths that increases in strength relative to the first
component with time. These two components do not necessarily arise due to
discrete ejection mechanisms, but may reflect that different parts of the
spectrum probe different physical depths and thus physical conditions,
through the wavelength-dependent expansion
opacity\cite{Karp1977,Barnes2013b}. Here we focus only on the thermal
component in the blue part of the spectrum and model it as a
blackbody with an extended envelope. We model the second component with
Gaussian emission lines in order not to bias the overall continuum fit at
shorter wavelengths, but do not interpret them. However, these features
clearly provide important information on the composition of the plasma and
must be addressed in future studies.

The expansion velocity of the gas can be inferred from the expansion of the
blackbody from the time of the explosion. Due to the optical thickness of a
blackbody, we would only be presented with the front face of the explosion.
Consequently, pure absorption features in the spectrum should be blueshifted by
the mean Doppler shift induced by the expansion speed of the gas.
Conservatively, we allow 0.1--0.3$c$ as the range of the
blueshift\cite{2017Natur.551...67P,2017arXiv171005432S,2017arXiv171109638W},
which depends on the details of the geometry of the system and thus restrict our
search for lines in the first epoch to rest wavelengths of 350\,nm and 810\,nm
$\times$ 1.1--1.3.

At the densities of the ejecta, the dominant source of opacity is expansion
opacity\cite{Kasen2013,2017Natur.551...80K}. This effect is able to
establish an apparent thermalisation through photo-equilibration of the
states\cite{2000ApJ...530..757P}. With wavelength-dependent opacity, the
physical depth traced at each wavelength varies. Because the large majority
of lines are in the blue end of the spectrum, the expansion opacity there
will be higher and, conversely, the physical depth shallower. This causes
the  relative strength of UV/near-infrared lines to change compared to the
pure LTE transmission values, with bluer absorption lines less prominent
relative to near-infrared ones. Additionally, because the population of
states is photo-equilibrated, metastable states will be enhanced relative
to non-metastable, as compared to LTE values\cite{1990sjws.conf..149J}.
It is therefore impossible, primarily due to the strongly
wavelength-dependent opacity, to use a simple comparison of LTE line-strengths across very different wavelengths.
Instead, we use independent optical depth parameters ($\tau$) for the two
absorption feature fits here. We also use the TARDIS code (see below) to achieve
a more self-consistent treatment with moving atmosphere, line-expansion opacity,
which shows the simultaneous presence of the $\sim0.4$ and $1\,\mu$m
Sr\,\textsc{ii} features.

\subsection{P\,Cygni modelling} The expansion velocity of the photosphere is
very high (0.2--0.3$c$).  At the measured temperature of the photosphere,
the thermal widths of individual lines are very narrow compared to the gross
velocity structure.  This means that the resonance region is very small and
the Sobolev approximation can be used in the Elementary Supernova (ES) model
as a prescription for the absorption structure near isolated lines\cite{1990sjws.conf..149J}.  We use the implementation of the P\,Cygni
profile in the ES from
\href{https://github.com/unoebauer/public-astro-tools}{https://github.com/unoebauer/public-astro-tools},
where the profile is parametrized in terms of the rest wavelength,
$\lambda_0$, the optical depth of the line, $\tau$, two scaling velocities
for the radial dependence of $\tau$, the photospheric velocity, and the
maximal velocity of the ejecta.  The latter two parameters specify the
velocity stratification.  The expansion velocity of the photosphere is
simultaneously used for the relativistic Doppler correction to the blackbody
temperature.  Additionally, because the implementation of the P\,Cygni
profile we are using does not include the relative population of the states
in the transition, we have included a parameter for enhancement/suppression
of the P\,Cygni emission component.

For practical reasons, we cannot fit all lines
simultaneously. However, fortunately, a handful of lines provide most of the
opacity. Because the relative opacity dictates the apparent strengths of the
lines, we divide the spectrum into ultraviolet/blue and red/infrared regions
to find the lines that will be strongest in their respective spectral region.
We do this because the opacity changes so severely from the infrared to the
optical (Fig.~\ref{fig:model_spectrum}). We make the division at 600\,nm where
the opacity increases sharply, however choosing 550\,nm or 700\,nm makes
no difference. We then include the strongest lines in each region (all lines
with a minimum strength of 20\% of the strongest line). The resultant lines are
the strong resonance lines from the ground state of Sr\,\textsc{ii} at
407.771 and 421.552\,nm, and the lines from the Sr\,\textsc{ii} 4p$^6$4d
metastable states at 1,032.731, 1,091.489, and 1,003.665\,nm.
These lines are all modelled using the same  P\,Cygni profile
prescription, where the relative strengths of each of the lines in the two
absorption complexes are set by the LTE relations, and in spite of the
relative simplicity of the analysis, provides a surprisingly good fit to the
data.

The final model we use to fit the spectrum is a relativistically-corrected
blackbody photosphere absorbed by an expanding atmosphere, containing the
five above-mentioned Sr\,\textsc{ii} transitions,
described by independent optical depths for the infrared
and ultraviolet lines.  The ratios of the lines
internally in each set are defined by their LTE strengths.  In the fitting
model we also use two additional Gaussian emission lines at long wavelengths
from the second emission component in order not to bias the long wavelength
continuum fit.  The best fit parameters and their associated errors are
found by sampling the posterior probability distributions of the parameters,
assuming flat priors on all parameters.  The fitting framework used is
\texttt{LMFIT}\cite{2016ascl.soft06014N} and the sampling is done using
\texttt{emcee}\cite{2013PASP..125..306F}.  We initiate 100 samplers, each
sampling for 1,000 steps.  We discard the first 100 steps as a burn-in phase
of the MCMC chains.  We use the median of the marginalized posterior
probability distribution as the best-fit values, and the 16th and 84th
percentiles as the uncertainties.  The best-fit models are shown in
Fig.~\ref{fig:pcygni_stack}.  The objective function, being highly
non-linear, causes the posterior probability distributions to be highly
complex and thus the
best-fit values difficult to optimize.  However, the
peak of the distributions are well centered, meaning the best-fit values are
well constrained, regardless of the posterior probability distribution
complexity.

\subsection{Expansion velocity evolution}

The fits constrain two independent parameters that can be used to infer the
velocity of the ejected material.  The photospheric expansion velocity used
to determine the width of the P\,Cygni line profile and the blackbody
radius, which scales with the square root of the observed luminosity and can
be converted to an expansion velocity based on the time of observation.
These two parameters are uncorrelated, as supported by the MCMC posterior
probability function samples, and therefore constitute two independent
measurements of the same physical quantity.  We show a plot of the evolution
of these two parameters in Extended Data Fig.~\ref{exfig:beta_evo}.  The
correspondence between the two estimates of the expansion velocity is
striking, especially given that the ratio of the estimates is
geometry-dependent, and we have assumed only simple spherical symmetry here.
Only the first epoch shows a somewhat discrepant
value, and there we expect a P\,Cygni model not to be entirely
applicable.  This close correspondence between the two independent measures
and the reasonable values inferred further supports the validity of the line
identification and the overall model.

\subsection{TARDIS modelling}

TARDIS\cite{Kerzendorf2014} is a Monte Carlo radiative-transfer spectral
synthesis code, where photons are essentially propagated through an expanding
atmosphere. Each photon will at any point have a probability of being
absorbed by an atomic transition, based on the wavelength of the photon,
strength of the line, and density of atomic species and electron population. A
synthetic spectrum can then be constructed by collecting the emergent photons.

To generate the synthetic spectra with TARDIS, we set up the physical models
using the inferred photospheric expansion velocities at the observed epochs. For
homologously expanding ejecta, the velocity of the atmosphere layers are at all
times specified by the outer edge expansion and the photospheric expansion. We
use the measured photospheric expansion velocity as the inner expansion velocity
and select the outer atmospheric velocity such that the bluest edge of the
developed absorption profiles in the synthetic spectra match the observed ones.
Currently, TARDIS only supports spherically symmetric explosions, so for
simplicity, we adopt this geometry. The kilonova ejecta are in most cases
likely to be asymmetric, due to the preferential motion of the mass in the plane of the orbit
of the two neutron stars. The neglect of deviation from spherical symmetry most
likely affects the absorption profiles and the inferred mass in the atmosphere,
as we could potentially only be seeing ejecta in a cone. Additionally, TARDIS
assumes a single photospheric velocity across the entire wavelength range. Due
to the strong wavelength dependence of the opacity, as discussed earlier, the
depth at which the photons escapes varies across the spectral coverage. Therefore,
the same reservations about inferring the mass in a given shell at a
given wavelength applies to the TARDIS simulations. This can be seen in effect
when choosing an ejecta density that matches the absorption feature at $350$\,nm,
because then the strength of the $810$\,nm absorption feature is significantly
overpredicted. Conversely, selecting an ejecta density that matches the
$810$\,nm absorption feature underpredicts the strength of the $350$\,nm
absorption.

At each epoch, the temperature of the photosphere is chosen so that an
atmosphere with no lines returns a blackbody-like spectrum which is similar to
the best-fit blackbody found in simple P\,Cygni model fits. Both the excitation
and ionization structure of the elements in the atmosphere are set
according to LTE where we assume for simplicity a constant temperature throughout
the atmosphere. This approach does not capture optical pumping of
metastable states and other non-LTE effects that will change the population of
the upper levels.

For the input abundances, we use the solar \emph{r}-process abundance ratio as
shown in Fig.~\ref{fig:r-process_abundances}, starting from $_{31}$Ga.  We run
the simulation in three steps, consecutively including heavier elements.  For
the first set of simulations, we include only the elements from $_{31}$Ga to
$_{37}$Rb and, as can be seen in Fig.~\ref{fig:r-process_abundances}, no lines
cause a significant deviation from a pure blackbody.  Next we include $_{38}$Sr
which forms the strong feature observed centred at $810$\,nm in the first epoch,
almost exclusively due to the three strong Sr\textsc{ii} lines at $\sim1\mu$m.
Last we run the same simulation, including all elements from $_{31}$Ga to
$_{92}$U. The feature at $810$\,nm is unaffected by the inclusion of the heavier
elements.

For the density, we initially adopt a power-law density structure of the ejecta,
parametrized in terms of velocity and epoch: $\rho(v,t)=\rho_0(t_0)^3
(v/v_0)^{n}$. We find that the line shapes depend on the assumed slope, where
for steeper slopes, a larger fraction of the line absorption is closer to the
line centre. We specify a density profile of $-3$, as is used in
ref.~\cite{2017PASJ...69..102T}, as this supported by the theoretical models and seem to
reproduce the absorption profiles relatively well. As also investigated in\cite{2017Natur.551...80K}, there is some freedom in the choice of slope, as it
is not well constrained from a modelling perspective and could have different
values depending on the matter ejection mechanism.

Adopting a single $\rho_0$ across all four epochs, with a $n = -3$, does not
yield synthetic spectra that match the observed spectra well around the
$810$\,nm $_{38}$Sr absorption feature across the epochs. If $\rho_0$ is chosen
to reproduce the strength of the $_{38}$Sr absorption feature of the first
epoch, the strength of the absorption feature is significantly overpredicted in
the later epochs using the same composition and assuming homology; the ejecta density has to be
scaled down by a factor of 5 in the subsequent epochs to match the spectrum. In other words, the
observed mass of Sr in the optically thin part of the spectrum inferred from the
TARDIS model for the first epoch spectrum appears to be significantly larger
than for the later epochs. Specifically, atmosphere masses of
$5\times 10^{-5}\,M_\odot$,
$1\times 10^{-5}\,M_\odot$, $1.2\times 10^{-5}\,M_\odot$, and
$1.3\times 10^{-5}\,M_\odot$ of $_{38}$Sr are required to reproduce the
observed absorption feature at $810$\,nm for the first four epochs respectively.

These numbers should be treated with some caution as this is the derived mass assuming spherical
symmetry, a fixed photospheric velocity, and no correction for light travel time
effects. They must be interpreted as lower limits to the total amount of
material ejected, as they only trace the matter between the photospheric front
and the outer atmosphere. Using the assumed solar abundances, these masses
correspond to this atmosphere having approximately 1\% of the total ejecta mass
inferred from lightcurve modelling\cite{2017Natur.551...80K}.

The TARDIS models additionally constrain the amount of the heavier \emph{r}-process
elements present in the outer, transparent layers of the ejecta. Using the
solar \emph{r}-process abundances with the inclusion of the heaviest elements,
the TARDIS synthetic spectra exhibit almost continuous absorption up to $\sim
6,000~\mathrm{\AA}$, which is not seen in the observed spectra. This point was also
touched upon earlier. The exact limit on the amount of heavy \emph{r}-process
material in the outer layers is difficult to infer accurately, based on the
simple models used, but our modelling indicates that the ratio of heavy
to light element abundance in this layer is significantly
lower than the solar \emph{r}-process ratio. This conclusion is consistent with the
inference made by other authors on the basis of the early blue colour of the
continuum spectrum\cite{McCully2017,Chornock2017}.

The inability of a single composition and density to reproduce the spectra
across the first four epochs may hint at a change in the elemental
abundance ratios as the photosphere recedes further into the ejecta.

The TARDIS models demonstrate that an isolated feature observed at $810$\,nm can
be produced by Sr and that no other known lines form this feature.
Additionally, the models hint at a possible variation in the abundances as the
deeper layers of the ejecta component are exposed, in line with what is
suggested by some models of NS mergers\cite{2014MNRAS.443.3134P}.

\subsection{Exclusion of the Cs\,\textsc{i} and Te\,\textsc{i} identification}
The Cs\,I 6s--6p resonance transitions\cite{2017Natur.551...75S} would of course
require Cs\,I to be present in the gas. But because Cs has the lowest first
ionisation potential of any element, the singly-charged ions of other
elements inevitably synthesised with Cs\cite{BBFH1957}, such as La\,\textsc{ii},
Eu\,\textsc{ii}, and Gd\,\textsc{ii}, are millions of times more abundant
than Cs\,\textsc{i} in an LTE plasma at close to the observed blackbody
temperatures. This problem is even worse at temperatures that produce
significant strong lines from Te\,\textsc{i}. These other elements will cause
absorption lines
that are at least two orders of magnitude stronger in the same wavelength
region as the proposed Cs and Te lines, e.g.\ the 706.62\,nm, 742.66\,nm,
or 929.05\,nm lines of La\,\textsc{ii}, Eu\,\textsc{ii}, and
Gd\,\textsc{ii} respectively, to name one of each. The same argument holds
for the excited state transition of Te\,\textsc{i} which has a very high
excitation energy of 5.49\,eV; the relative population of the
Te\,\textsc{i} excited state is extremely low, less than $10^{-7}$. Thus,
no realistic scenario exists in which either of these lines can be detected
without orders of magnitude stronger lines from other elements dominating.

\end{methods}

\clearpage

\begin{addendum}
 \item[Acknowledgements]
	We thank Masaomi Tanaka for revisiting his previous analysis for us and for
  access to his spectra and line lists. We thank Jens Hjorth and Nanda Rea for
  useful discussions.
	We thank the ESO Director General for allocating Discretionary Time to this programme
	and the ESO operation staff for excellent support.
	DW, DBM, and JS are supported in part by Independent Research Fund Denmark grant DFF~-~7014-00017.
	The Cosmic Dawn Center is funded by the Danish National Research Foundation.
  AA is supported by the European Research Council through ERC Starting Grant No.~677912 EUROPIUM.
	AB is supported by the European Research Council through ERC Starting
	Grant No.~759253 GreatMoves and by the Sonderforschungsbereich
  SFB 881 ``The Milky Way System'' (subproject A10) of the German Research
  Foundation (DFG) and acknowledges support by the Klaus Tschira Foundation.
	SC acknowledges partial funding from Agenzia Spaziale Italiana-Istituto Nazionale di Astrofisica grant I/004/11/3.
	GL is supported by a research grant (19054) from Villum Fonden.
	KEH acknowledges support by a Project Grant (162948--051) from The Icelandic Research Fund.
  AJL acknowledges funding from the European Research Council under grant agreement No.~725246,
	and from STFC via grant number ST/P000495/1.
	EP acknowledges funding from ASI INAF grant I/088/06/0,
	and from INAF project:
	`Gravitational Wave Astronomy with the first detections of aLIGO and aVIRGO experiments'

\item[Author contributions]

DW, CJH, and JS were the primary drivers of the project;
AK, DBM, JPUF, and ACA were involved in important discussions developing the understanding of the physical processes.
All authors contributed to discussions and editing the paper.
DW did
the initial blackbody with absorber fits to the first epoch spectrum
and created Figs.~\ref{fig:first_epoch_spectrum}
and \ref{fig:model_spectrum},
and Extended Data Figs.~\ref{exfig:LTE_individual_spectra} and \ref{exfig:expansion_opacity_plot},
made the initial line identification,
recognised the P\,Cygni profiles in the later epochs,
wrote the LTE code,
and was the primary author of the main text.
CJH computed the initial models and synthetic spectra with MOOG,
and generated the MOOG spectra for HD\,88609 and CS\,22892$-$052.
CJH and AK produced the MOOG spectra from 3000--20\,000\,\AA\ for the kilonova template photosphere for all heavy elements.
CJH wrote related sections on MOOG spectrum synthesis
and significant parts of the text on nucleosynthesis.
AK provided the line lists.
JS reduced and processed all the X-shooter data,
produced the P\,Cygni fitting codes
and fit the P\,Cygni profiles to all epochs, as well as extending the TARDIS
code to include the Kurucz linelists and implemented the TARDIS modelling.
JS also produced Fig.~\ref{fig:pcygni_stack}
and Extended Data Figs.~\ref{exfig:moog_model_spectrum},
\ref{exfig:moog_individual_spectra_plot}, and
\ref{exfig:tardis_model},
wrote the related methods sections
and a significant part of the main text.
 
\item[Author Information] Reprints and permissions information is available at \url{www.nature.com/reprints}.
The authors declare no
competing financial interests. Correspondence and requests for materials should be
addressed to D.~W. (darach@nbi.ku.dk).

\item[Data Availability Statement]
Work in this paper was based on observations made with ESO Telescopes at the Paranal Observatory
under programmes
099.D-0382 (PI: E.~Pian),
099.D-0622 (PI: P.~D'Avanzo),
099.D-0376 (PI: S.~J.~Smartt), and
099.D-0191 (PI: A.~Grado).
 
\end{addendum}
 
\clearpage

\begin{table}
 \includegraphics[width=\columnwidth,clip=]{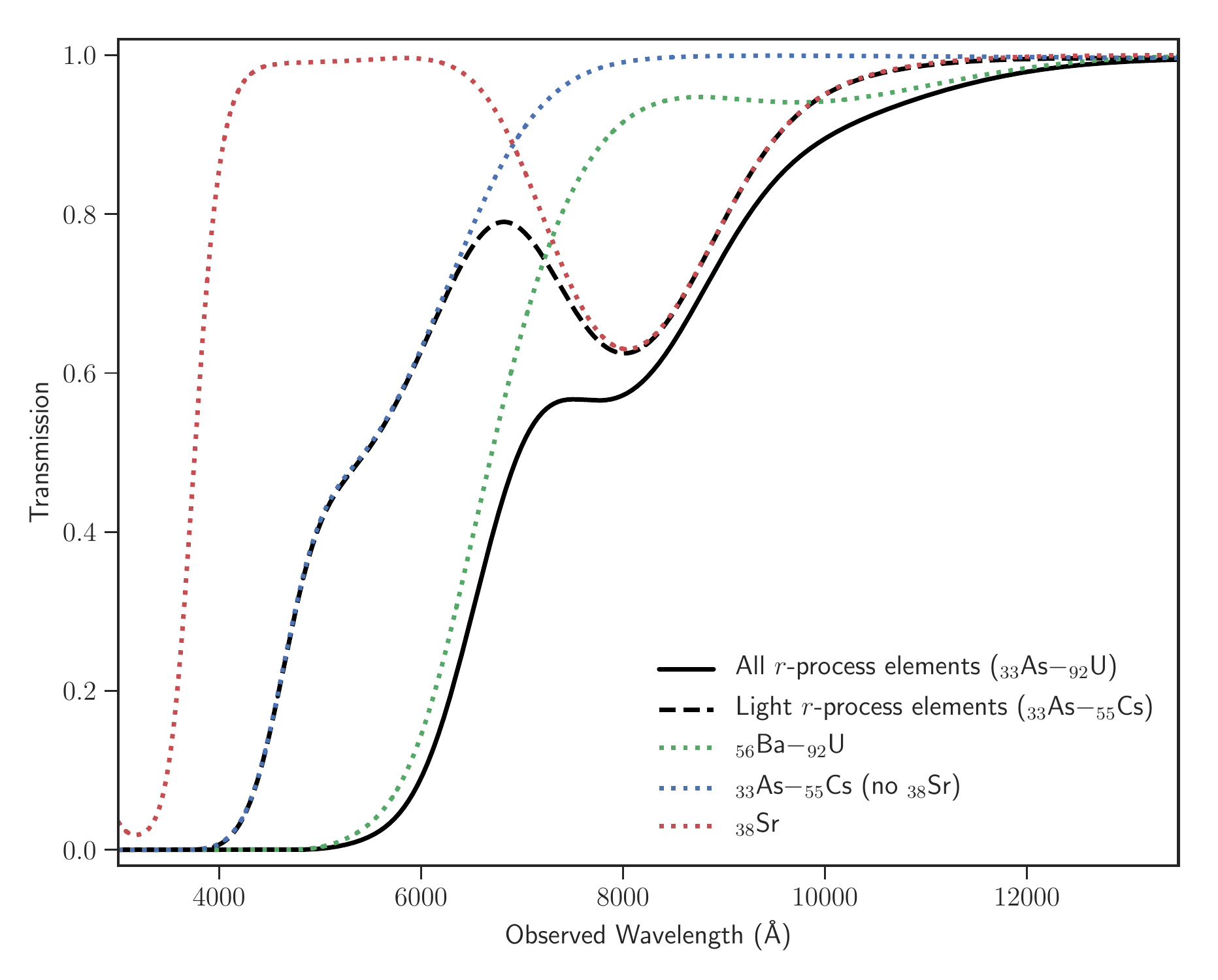}
 \caption{\textbf{Synthetic \emph{r}-process element transmission spectra.} The spectra
          are generated using MOOG,
          where the relative abundances are based on solar
          \emph{r}-process abundances. The spectra are blueshifted,
          broadened and normalized as in Fig.~\ref{fig:model_spectrum}. The
          solid, black line is the total transmission spectrum for an
          atmosphere containing all the \emph{r}-process elements
          ($_{33}$As$-_{92}$U). The dashed, black line is the same spectrum,
          only including the light \emph{r}-process elements
          ($_{33}$As$-_{55}$Cs). The contributions from different subsets of
          species are also shown. The green, dotted line shows the
          heavy $r$-process elements ($_{56}$Ba$-_{92}$U), the blue, dotted
          lines shows the light \emph{r}-process elements ($_{33}$As$-_{55}$Cs
          excluding Sr), which are both shown individually as thin,
          black lines and summed in the red, dotted line. This plot shows
          how Sr stands out in absorption, regardless of the
          composition of the material. The normalization is arbitrary
          and different to the LTE equivalent in
          Fig.~\ref{fig:model_spectrum} for display reasons.
          \label{exfig:moog_model_spectrum}
         }
\end{table}

\begin{table}
 \includegraphics[width=\columnwidth,clip=]{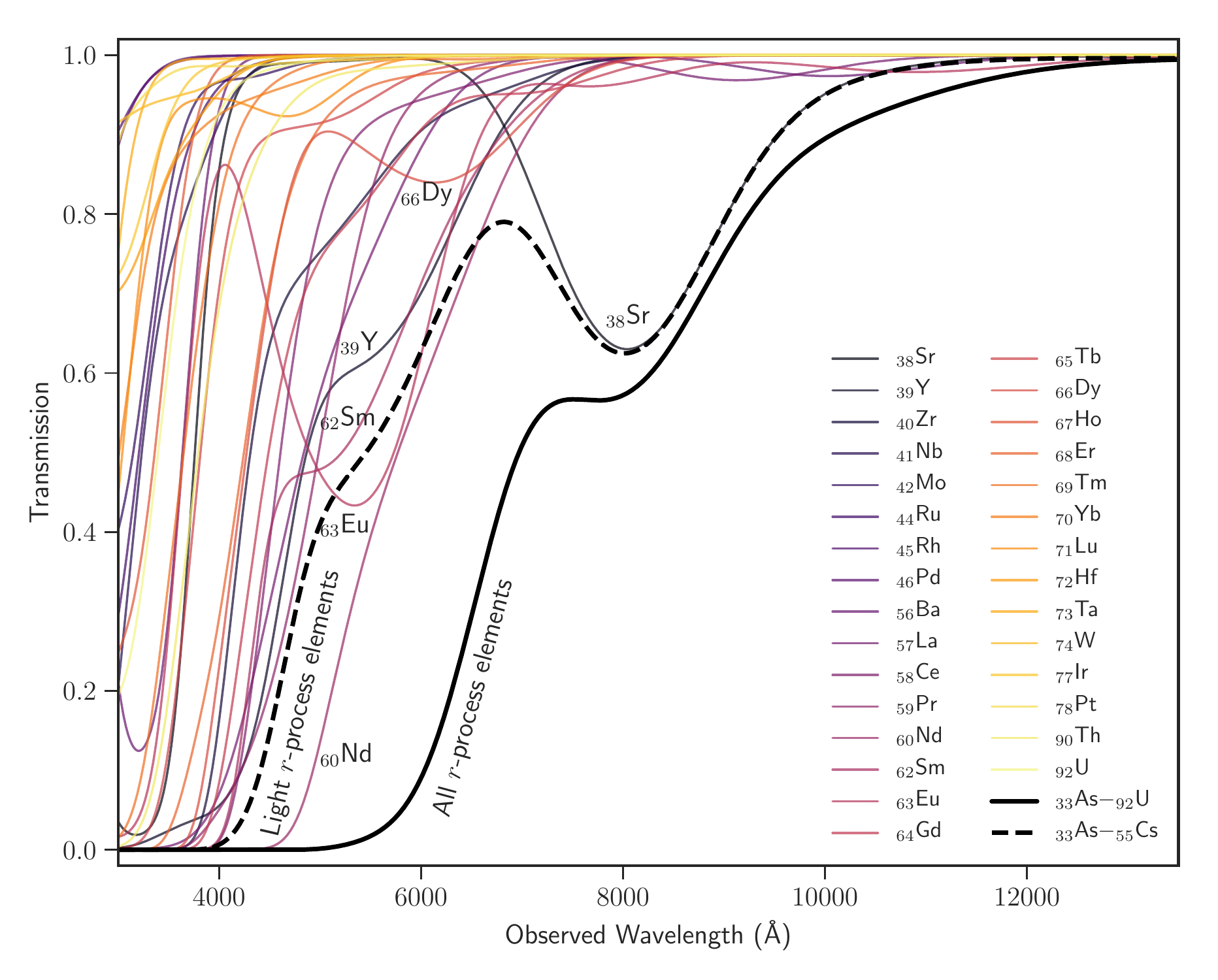}
 \caption{\textbf{Synthetic \emph{r}-process transmission spectra.}
          The spectra are generated with MOOG and are similar to Extended Data
          Fig.~\ref{exfig:moog_model_spectrum},
          except that all element contributions are shown individually. The
          elements contributing most at the reddest wavelengths are noted in
          the plot.
          \label{exfig:moog_individual_spectra_plot}
         }
\end{table}

\begin{table}
 \includegraphics[width=\columnwidth,clip=]{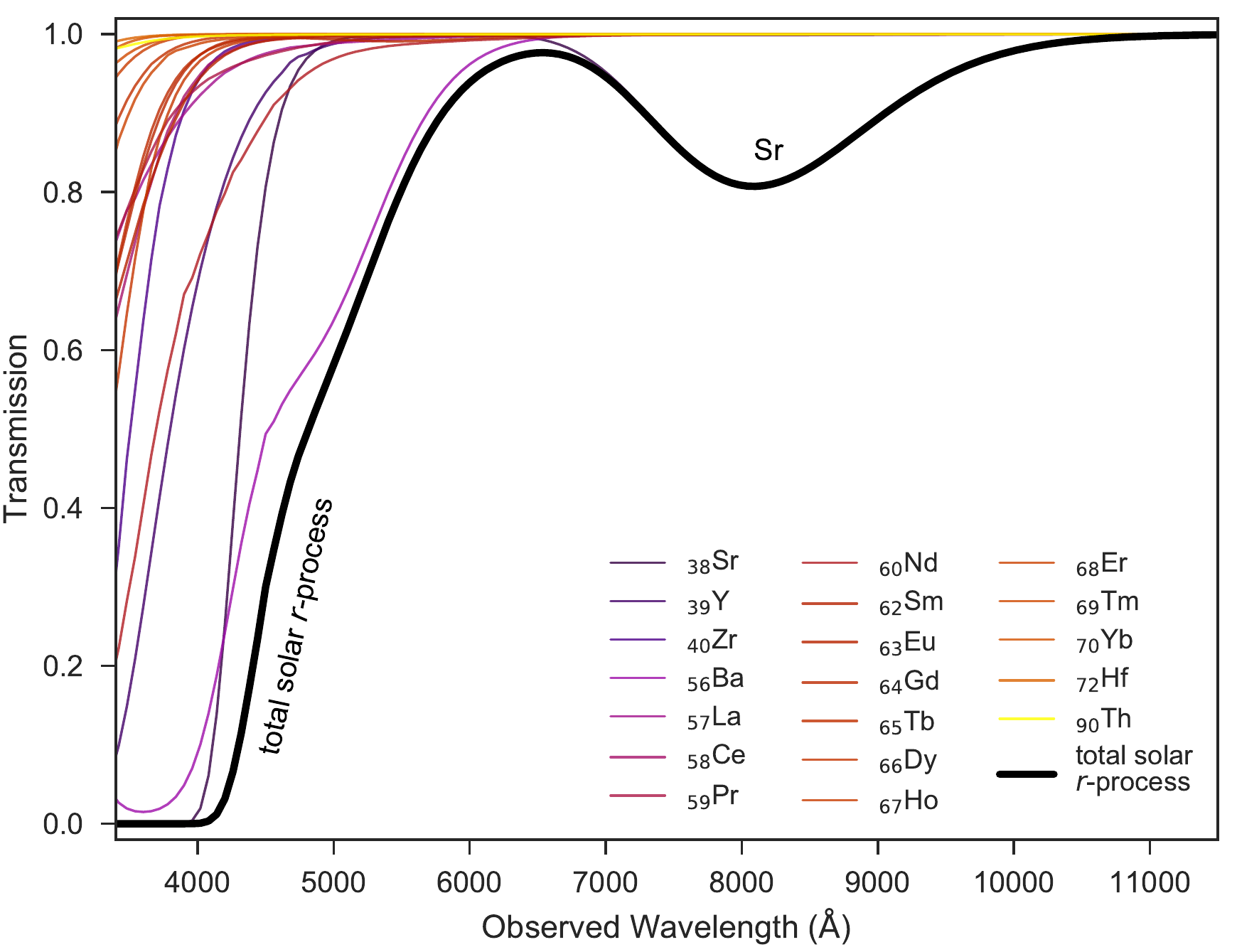}
 \caption{\textbf{Thermal transmission spectra for \emph{r}-process elements plotted
          individually.} The spectra are based on the lines formed in a gas in
          local thermal equilibrium. The abundances of elements are scaled to
          the solar
          \emph{r}-process and the spectra are velocity broadened, blueshifted
          and normalised as in Fig.~\ref{fig:model_spectrum}. The spectrum
          derived from the total solar \emph{r}-process abundance mix is plotted
          as a thick black line. The contributions from Sr clearly dominate at
          $\sim8,000$\,\AA, with no significant contribution from any other
          element.
          \label{exfig:LTE_individual_spectra}
         }
\end{table}

\begin{table}
 \includegraphics[width=\columnwidth,clip=]{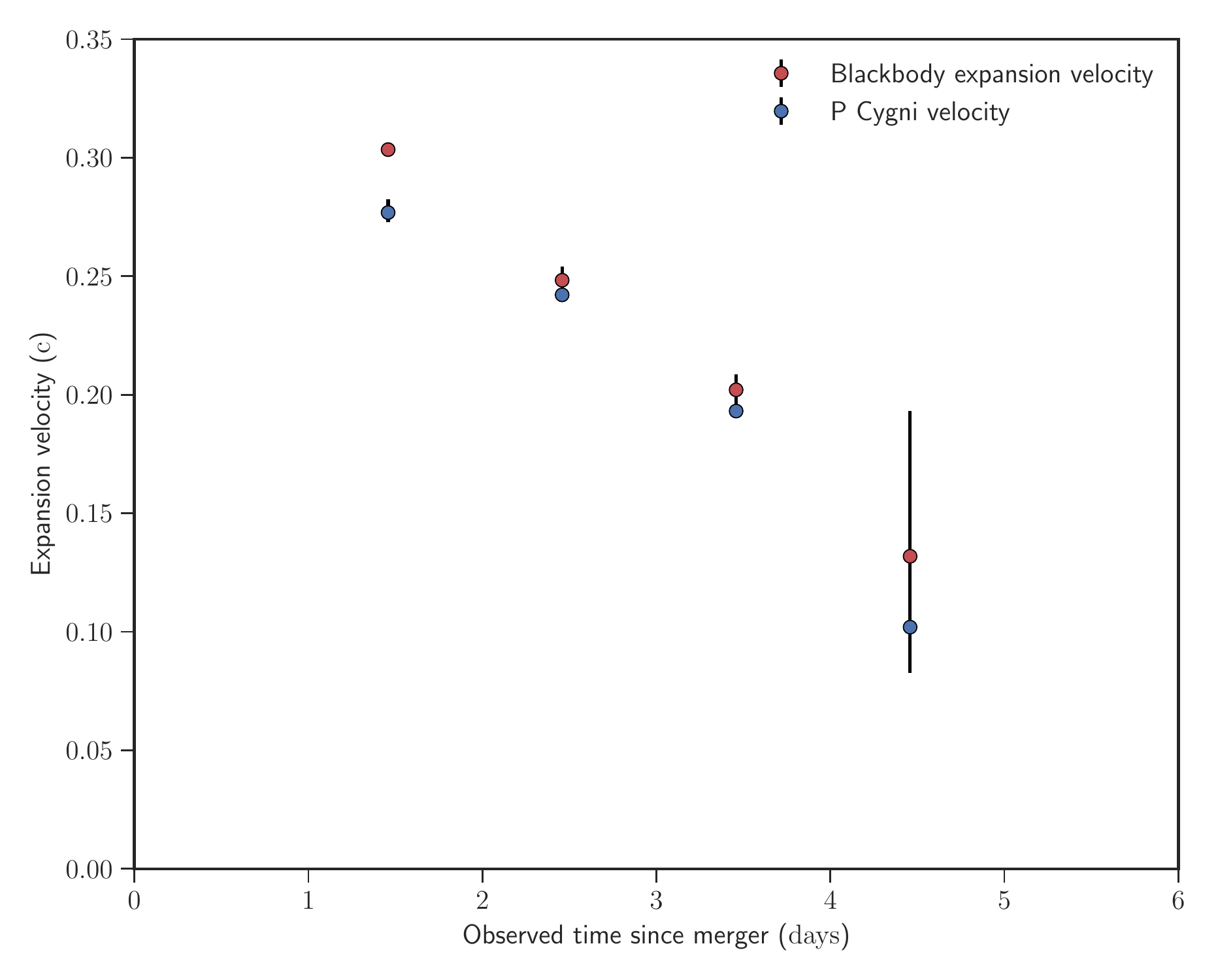}
 \caption{\textbf{Evolution of the ejecta expansion velocity.} The velocities are
          determined independently from the P\,Cygni absorption line widths
          (blue points) and the blackbody radius
          (red points). Uncertainties shown are $1\sigma$. The correspondence
          between the two independent estimates is striking.
          \label{exfig:beta_evo}
         }
\end{table}

\begin{table}
\includegraphics[width=\columnwidth,clip=]{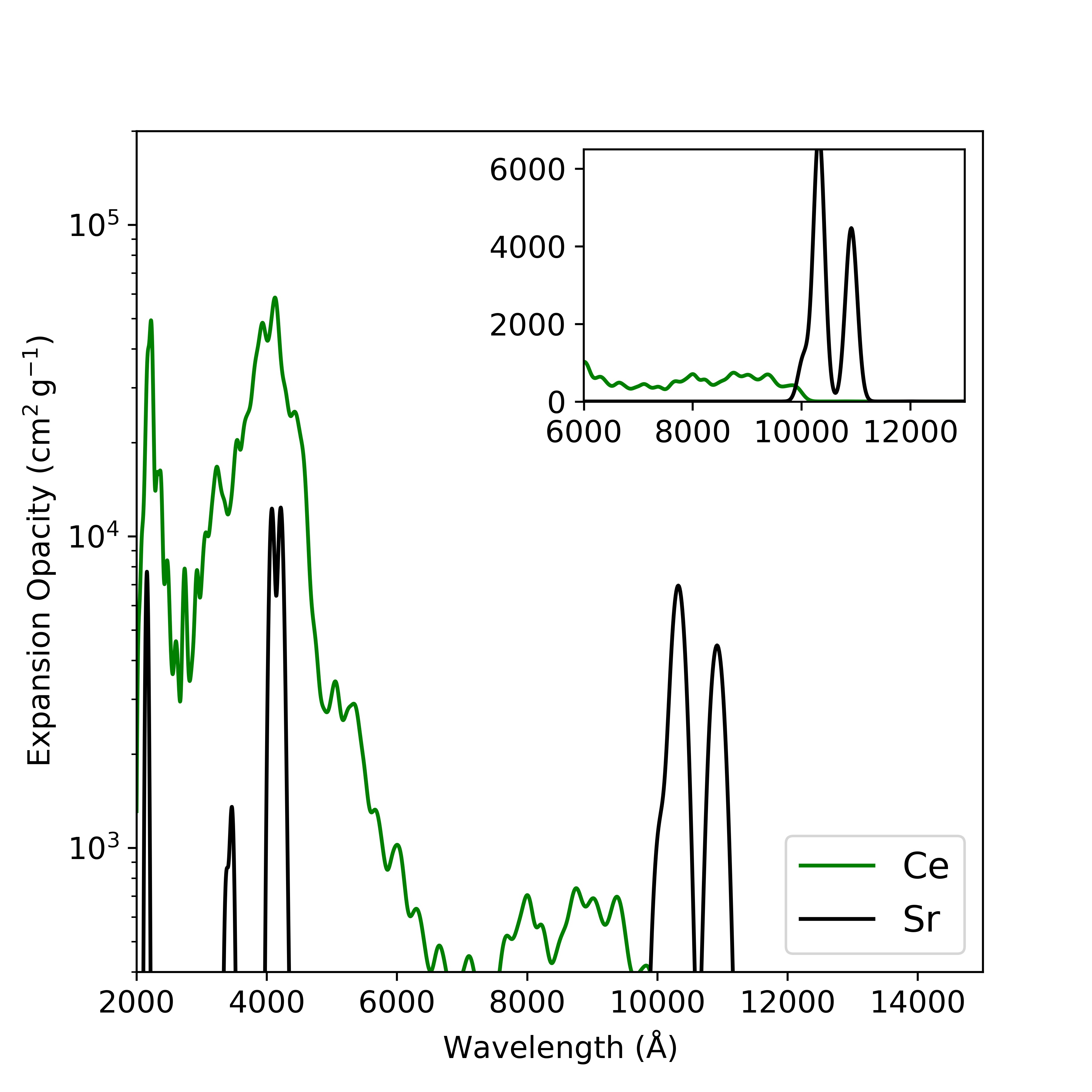}
\caption{\textbf{Comparison of the expansion opacities at modest optical depths for Sr
         and Ce.} This calculation shows the
         potential of Sr to dominate the opacity at $\sim1\,\mu$m at low optical
         depths. The opacities are based on local thermal
         equilibrium calculations for a gas at a temperature of 5,000\,K, a mean
         local density of $8.4\times10^{-17}$\,g\,cm$^{-3}$ of Sr or Ce, an
         electron density of $7.6\times10^{8}$\,cm$^{-3}$, and a 1\% atmospheric
         radius at 1.5\,days after the explosion. Line lists used for Sr and Ce
         are from the Kurucz and VALD databases respectively.
         \label{exfig:expansion_opacity_plot}
        }
\end{table}

\begin{table}
 \includegraphics[width=\columnwidth,clip=]{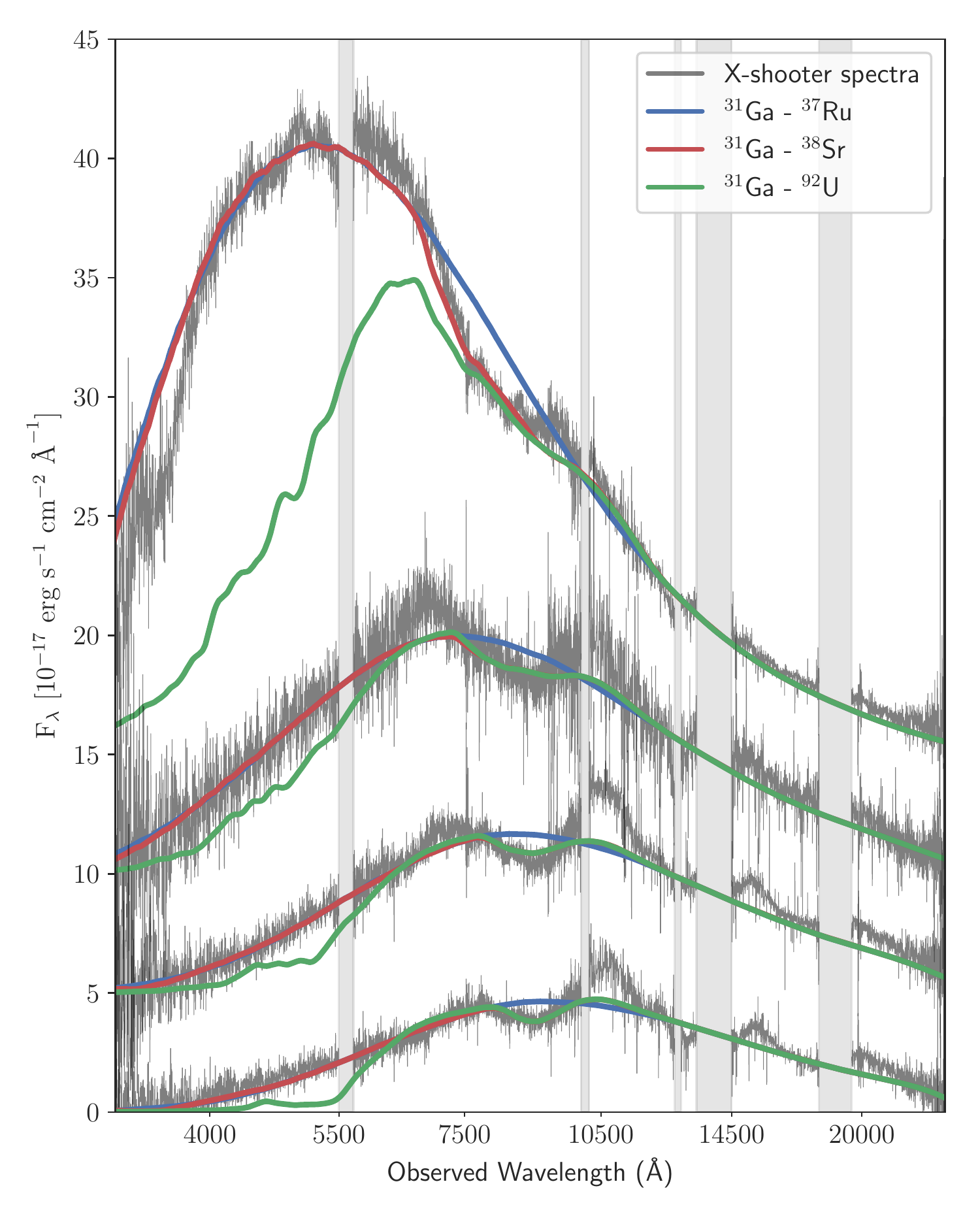}
 \caption{\textbf{Radiative transfer models
          from the first four epochs using the TARDIS code.} The blue line is
          the synthetic
          TARDIS spectrum using relative solar \emph{r}-process abundances
          and including elements from $_{31}$Ga to $_{37}$Rb, i.e.\ without
          Sr. The red line
          additionally includes $_{38}$Sr. The
          green line is a model including all elements from $_{31}$Ga to
          $_{92}$U. These models show that the spectra are well-reproduced
          with elements around the first \emph{r}-process abundance peak,
          specifically Sr.
          \label{exfig:tardis_model}
         }
\end{table}


\begin{thebibliography}{10}
\expandafter\ifx\csname url\endcsname\relax
  \def\url#1{\texttt{#1}}\fi
\expandafter\ifx\csname urlprefix\endcsname\relax\def\urlprefix{URL }\fi
\providecommand{\bibinfo}[2]{#2}

\bibitem{BBFH1957}
\bibinfo{author}{{Burbidge}, E.~M.}, \bibinfo{author}{{Burbidge}, G.~R.},
  \bibinfo{author}{{Fowler}, W.~A.} \& \bibinfo{author}{{Hoyle}, F.}
\newblock \bibinfo{title}{{Synthesis of the Elements in Stars}}.
\newblock \emph{\bibinfo{journal}{Rev. Mod. Phys.}}
  \textbf{\bibinfo{volume}{29}}, \bibinfo{pages}{547--650}
  (\bibinfo{year}{1957}).

\bibitem{2019Natur.569..241S}
\bibinfo{author}{{Siegel}, D.~M.}, \bibinfo{author}{{Barnes}, J.} \&
  \bibinfo{author}{{Metzger}, B.~D.}
\newblock \bibinfo{title}{{Collapsars as a major source of r-process
  elements}}.
\newblock \emph{\bibinfo{journal}{Nature}} \textbf{\bibinfo{volume}{569}},
  \bibinfo{pages}{241--244} (\bibinfo{year}{2019}).

\bibitem{Lattimer1977}
\bibinfo{author}{{Lattimer}, J.~M.}, \bibinfo{author}{{Mackie}, F.},
  \bibinfo{author}{{Ravenhall}, D.~G.} \& \bibinfo{author}{{Schramm}, D.~N.}
\newblock \bibinfo{title}{{The decompression of cold neutron star matter}}.
\newblock \emph{\bibinfo{journal}{Astrophys. J.}}
  \textbf{\bibinfo{volume}{213}}, \bibinfo{pages}{225--233}
  (\bibinfo{year}{1977}).

\bibitem{Eichler1989}
\bibinfo{author}{{Eichler}, D.}, \bibinfo{author}{{Livio}, M.},
  \bibinfo{author}{{Piran}, T.} \& \bibinfo{author}{{Schramm}, D.~N.}
\newblock \bibinfo{title}{{Nucleosynthesis, neutrino bursts and gamma-rays from
  coalescing neutron stars}}.
\newblock \emph{\bibinfo{journal}{Nature}} \textbf{\bibinfo{volume}{340}},
  \bibinfo{pages}{126--128} (\bibinfo{year}{1989}).

\bibitem{Freiburghaus1999}
\bibinfo{author}{{Freiburghaus}, C.}, \bibinfo{author}{{Rosswog}, S.} \&
  \bibinfo{author}{{Thielemann}, F.-K.}
\newblock \bibinfo{title}{{R-Process in Neutron Star Mergers}}.
\newblock \emph{\bibinfo{journal}{Astrophys. J. Lett.}}
  \textbf{\bibinfo{volume}{525}}, \bibinfo{pages}{L121--L124}
  (\bibinfo{year}{1999}).

\bibitem{Ji2016}
\bibinfo{author}{{Ji}, A.~P.}, \bibinfo{author}{{Frebel}, A.},
  \bibinfo{author}{{Simon}, J.~D.} \& \bibinfo{author}{{Chiti}, A.}
\newblock \bibinfo{title}{{Complete Element Abundances of Nine Stars in the
  r-process Galaxy Reticulum II}}.
\newblock \emph{\bibinfo{journal}{Astrophys. J.}}
  \textbf{\bibinfo{volume}{830}}, \bibinfo{pages}{93} (\bibinfo{year}{2016}).

\bibitem{Metzger2010}
\bibinfo{author}{{Metzger}, B.~D.} \emph{et~al.}
\newblock \bibinfo{title}{{Electromagnetic counterparts of compact object
  mergers powered by the radioactive decay of r-process nuclei}}.
\newblock \emph{\bibinfo{journal}{Mon. Not. R. Astron. Soc.}}
  \textbf{\bibinfo{volume}{406}}, \bibinfo{pages}{2650--2662}
  (\bibinfo{year}{2010}).

\bibitem{Barnes&Kasen2013}
\bibinfo{author}{{Barnes}, J.} \& \bibinfo{author}{{Kasen}, D.}
\newblock \bibinfo{title}{{Effect of a High Opacity on the Light Curves of
  Radioactively Powered Transients from Compact Object Mergers}}.
\newblock \emph{\bibinfo{journal}{Astrophys. J.}}
  \textbf{\bibinfo{volume}{775}}, \bibinfo{pages}{18} (\bibinfo{year}{2013}).

\bibitem{Tanvir2013}
\bibinfo{author}{{Tanvir}, N.~R.} \emph{et~al.}
\newblock \bibinfo{title}{{A `kilonova' associated with the short-duration
  {$\gamma$}-ray burst GRB 130603B}}.
\newblock \emph{\bibinfo{journal}{Nature}} \textbf{\bibinfo{volume}{500}},
  \bibinfo{pages}{547--549} (\bibinfo{year}{2013}).

\bibitem{2017PhRvL.119p1101A}
\bibinfo{author}{{Abbott}, B.~P.}, \bibinfo{author}{{Abbott}, R.} \&
  \bibinfo{author}{et~al.}
\newblock \bibinfo{title}{{GW170817: Observation of Gravitational Waves from a
  Binary Neutron Star Inspiral}}.
\newblock \emph{\bibinfo{journal}{Physical Review Letters}}
  \textbf{\bibinfo{volume}{119}}, \bibinfo{pages}{161101}
  (\bibinfo{year}{2017}).

\bibitem{2017Natur.551...67P}
\bibinfo{author}{{Pian}, E.} \emph{et~al.}
\newblock \bibinfo{title}{{Spectroscopic identification of r-process
  nucleosynthesis in a double neutron-star merger}}.
\newblock \emph{\bibinfo{journal}{Nature}} \textbf{\bibinfo{volume}{551}},
  \bibinfo{pages}{67--70} (\bibinfo{year}{2017}).

\bibitem{2017Natur.551...75S}
\bibinfo{author}{{Smartt}, S.~J.} \emph{et~al.}
\newblock \bibinfo{title}{{A kilonova as the electromagnetic counterpart to a
  gravitational-wave source}}.
\newblock \emph{\bibinfo{journal}{Nature}} \textbf{\bibinfo{volume}{551}},
  \bibinfo{pages}{75--79} (\bibinfo{year}{2017}).

\bibitem{Baade&Zwicky1934}
\bibinfo{author}{{Baade}, W.} \& \bibinfo{author}{{Zwicky}, F.}
\newblock \bibinfo{title}{{Cosmic Rays from Super-novae}}.
\newblock \emph{\bibinfo{journal}{Proceedings of the National Academy of
  Science}} \textbf{\bibinfo{volume}{20}}, \bibinfo{pages}{259--263}
  (\bibinfo{year}{1934}).

\bibitem{Tanaka2013}
\bibinfo{author}{{Tanaka}, M.} \& \bibinfo{author}{{Hotokezaka}, K.}
\newblock \bibinfo{title}{{Radiative Transfer Simulations of Neutron Star
  Merger Ejecta}}.
\newblock \emph{\bibinfo{journal}{Astrophys. J.}}
  \textbf{\bibinfo{volume}{775}}, \bibinfo{pages}{113} (\bibinfo{year}{2013}).

\bibitem{2017Natur.551...80K}
\bibinfo{author}{{Kasen}, D.}, \bibinfo{author}{{Metzger}, B.},
  \bibinfo{author}{{Barnes}, J.}, \bibinfo{author}{{Quataert}, E.} \&
  \bibinfo{author}{{Ramirez-Ruiz}, E.}
\newblock \bibinfo{title}{{Origin of the heavy elements in binary neutron-star
  mergers from a gravitational-wave event}}.
\newblock \emph{\bibinfo{journal}{Nature}} \textbf{\bibinfo{volume}{551}},
  \bibinfo{pages}{80--84} (\bibinfo{year}{2017}).

\bibitem{2012ascl.soft02009S}
\bibinfo{author}{{Sneden}, C.}, \bibinfo{author}{{Bean}, J.},
  \bibinfo{author}{{Ivans}, I.}, \bibinfo{author}{{Lucatello}, S.} \&
  \bibinfo{author}{{Sobeck}, J.}
\newblock \bibinfo{title}{{MOOG: LTE line analysis and spectrum synthesis}}.
\newblock \bibinfo{howpublished}{Astrophysics Source Code Library}
  (\bibinfo{year}{2012}).

\bibitem{Kerzendorf2014}
\bibinfo{author}{{Kerzendorf}, W.~E.} \& \bibinfo{author}{{Sim}, S.~A.}
\newblock \bibinfo{title}{{A spectral synthesis code for rapid modelling of
  supernovae}}.
\newblock \emph{\bibinfo{journal}{Mon. Not. R. Astron. Soc.}}
  \textbf{\bibinfo{volume}{440}}, \bibinfo{pages}{387--404}
  (\bibinfo{year}{2014}).

\bibitem{Lodders2011}
\bibinfo{author}{{Lodders}, K.}, \bibinfo{author}{{Palme}, H.} \&
  \bibinfo{author}{{Gail}, H.-P.}
\newblock \bibinfo{title}{{Abundances of the Elements in the Solar System}}.
\newblock \emph{\bibinfo{journal}{Landolt B{\"o}rnstein}}
  (\bibinfo{year}{2009}).
\newblock \bibinfo{note}{2011 update arxiv:0901.1149v2}.

\bibitem{2014ApJ...787...10B}
\bibinfo{author}{{Bisterzo}, S.}, \bibinfo{author}{{Travaglio}, C.},
  \bibinfo{author}{{Gallino}, R.}, \bibinfo{author}{{Wiescher}, M.} \&
  \bibinfo{author}{{K{\"a}ppeler}, F.}
\newblock \bibinfo{title}{{Galactic Chemical Evolution and Solar s-process
  Abundances: Dependence on the $^{13}$C-pocket Structure}}.
\newblock \emph{\bibinfo{journal}{Astrophys. J.}}
  \textbf{\bibinfo{volume}{787}}, \bibinfo{pages}{10} (\bibinfo{year}{2014}).

\bibitem{Honda2007}
\bibinfo{author}{{Honda}, S.}, \bibinfo{author}{{Aoki}, W.},
  \bibinfo{author}{{Ishimaru}, Y.} \& \bibinfo{author}{{Wanajo}, S.}
\newblock \bibinfo{title}{{Neutron-Capture Elements in the Very Metal-poor Star
  HD 88609: Another Star with Excesses of Light Neutron-Capture Elements}}.
\newblock \emph{\bibinfo{journal}{Astrophys. J.}}
  \textbf{\bibinfo{volume}{666}}, \bibinfo{pages}{1189--1197}
  (\bibinfo{year}{2007}).

\bibitem{2000ApJ...533L.139S}
\bibinfo{author}{{Sneden}, C.} \emph{et~al.}
\newblock \bibinfo{title}{{Evidence of Multiple R-Process Sites in the Early
  Galaxy: New Observations of CS 22892-052}}.
\newblock \emph{\bibinfo{journal}{Astrophys. J. Lett.}}
  \textbf{\bibinfo{volume}{533}}, \bibinfo{pages}{L139--L142}
  (\bibinfo{year}{2000}).

\bibitem{Kasen2013}
\bibinfo{author}{Kasen, D.}, \bibinfo{author}{Badnell, N.~R.} \&
  \bibinfo{author}{Barnes, J.}
\newblock \bibinfo{title}{{Opacities and spectra of the r-process ejecta from
  neutron star mergers}}.
\newblock \emph{\bibinfo{journal}{Astrophys. J.}}
  \textbf{\bibinfo{volume}{774}}, \bibinfo{pages}{25} (\bibinfo{year}{2013}).

\bibitem{1990sjws.conf..149J}
\bibinfo{author}{{Jeffery}, D.~J.} \& \bibinfo{author}{{Branch}, D.}
\newblock \bibinfo{title}{{Analysis of Supernova Spectra}}.
\newblock In \bibinfo{editor}{{Wheeler}, J.~C.}, \bibinfo{editor}{{Piran}, T.}
  \& \bibinfo{editor}{{Weinberg}, S.} (eds.)
  \emph{\bibinfo{booktitle}{Supernovae, Jerusalem Winter School for Theoretical
  Physics}}, \bibinfo{pages}{149} (\bibinfo{year}{1990}).

\bibitem{Kurucz2017}
\bibinfo{author}{{Kurucz}, R.~L.}
\newblock \bibinfo{title}{{Including all the lines: data releases for spectra
  and opacities}}.
\newblock \emph{\bibinfo{journal}{Canadian Journal of Physics}}
  \textbf{\bibinfo{volume}{95}}, \bibinfo{pages}{825--827}
  (\bibinfo{year}{2017}).

\bibitem{Wanajo2014}
\bibinfo{author}{{Wanajo}, S.} \emph{et~al.}
\newblock \bibinfo{title}{{Production of All the r-process Nuclides in the
  Dynamical Ejecta of Neutron Star Mergers}}.
\newblock \emph{\bibinfo{journal}{Astrophys. J. Lett.}}
  \textbf{\bibinfo{volume}{789}}, \bibinfo{pages}{L39} (\bibinfo{year}{2014}).

\bibitem{Just2015}
\bibinfo{author}{{Just}, O.}, \bibinfo{author}{{Bauswein}, A.},
  \bibinfo{author}{{Pulpillo}, R.~A.}, \bibinfo{author}{{Goriely}, S.} \&
  \bibinfo{author}{{Janka}, H.-T.}
\newblock \bibinfo{title}{{Comprehensive nucleosynthesis analysis for ejecta of
  compact binary mergers}}.
\newblock \emph{\bibinfo{journal}{Mon. Not. R. Astron. Soc.}}
  \textbf{\bibinfo{volume}{448}}, \bibinfo{pages}{541--567}
  (\bibinfo{year}{2015}).

\bibitem{Drout2017}
\bibinfo{author}{{Drout}, M.~R.} \emph{et~al.}
\newblock \bibinfo{title}{{Light curves of the neutron star merger
  GW170817/SSS17a: Implications for r-process nucleosynthesis}}.
\newblock \emph{\bibinfo{journal}{Science}} \textbf{\bibinfo{volume}{358}},
  \bibinfo{pages}{1570--1574} (\bibinfo{year}{2017}).

\bibitem{2017ApJ...848L..27T}
\bibinfo{author}{{Tanvir}, N.~R.} \emph{et~al.}
\newblock \bibinfo{title}{{The Emergence of a Lanthanide-rich Kilonova
  Following the Merger of Two Neutron Stars}}.
\newblock \emph{\bibinfo{journal}{Astrophys. J. Lett.}}
  \textbf{\bibinfo{volume}{848}}, \bibinfo{pages}{L27} (\bibinfo{year}{2017}).

\bibitem{2014ApJ...797..123H}
\bibinfo{author}{{Hansen}, C.~J.}, \bibinfo{author}{{Montes}, F.} \&
  \bibinfo{author}{{Arcones}, A.}
\newblock \bibinfo{title}{{How Many Nucleosynthesis Processes Exist at Low
  Metallicity?}}
\newblock \emph{\bibinfo{journal}{Astrophys. J.}}
  \textbf{\bibinfo{volume}{797}}, \bibinfo{pages}{123} (\bibinfo{year}{2014}).

\bibitem{Hewish1968}
\bibinfo{author}{{Hewish}, A.}, \bibinfo{author}{{Bell}, S.~J.},
  \bibinfo{author}{{Pilkington}, J.~D.~H.}, \bibinfo{author}{{Scott}, P.~F.} \&
  \bibinfo{author}{{Collins}, R.~A.}
\newblock \bibinfo{title}{{Observation of a Rapidly Pulsating Radio Source}}.
\newblock \emph{\bibinfo{journal}{Nature}} \textbf{\bibinfo{volume}{217}},
  \bibinfo{pages}{709--713} (\bibinfo{year}{1968}).


\end{thebibliography}

\begin{thebibliography}{10}
\expandafter\ifx\csname url\endcsname\relax
  \def\url#1{\texttt{#1}}\fi
\expandafter\ifx\csname urlprefix\endcsname\relax\def\urlprefix{URL }\fi
\providecommand{\bibinfo}[2]{#2}

\setcounter{enumiv}{30}

\bibitem{MOOG_software}
\bibinfo{title}{{MOOG} spectral synthesis code}.
\newblock
  \bibinfo{howpublished}{\url{https://www.as.utexas.edu/~chris/moog.html}}.

\bibitem{2004astro.ph..5087C}
\bibinfo{author}{{Castelli}, F.} \& \bibinfo{author}{{Kurucz}, R.~L.}
\newblock \bibinfo{title}{{New Grids of ATLAS9 Model Atmospheres}}.
\newblock \emph{\bibinfo{journal}{arXiv Astrophysics e-prints}}
  (\bibinfo{year}{2004}).

\bibitem{Biemont2003lines}
\bibinfo{author}{{Bi{\'e}mont}, E.} \& \bibinfo{author}{{Quinet}, P.}
\newblock \bibinfo{title}{{Recent Advances in the Study of Lanthanide Atoms and
  Ions}}.
\newblock \emph{\bibinfo{journal}{Physica Scripta Volume T}}
  \textbf{\bibinfo{volume}{105}}, \bibinfo{pages}{38} (\bibinfo{year}{2003}).

\bibitem{DenHartog2003Nd}
\bibinfo{author}{{Den Hartog}, E.~A.}, \bibinfo{author}{{Lawler}, J.~E.},
  \bibinfo{author}{{Sneden}, C.} \& \bibinfo{author}{{Cowan}, J.~J.}
\newblock \bibinfo{title}{{Improved Laboratory Transition Probabilities for Nd
  II and Application to the Neodymium Abundances of the Sun and Three
  Metal-poor Stars}}.
\newblock \emph{\bibinfo{journal}{Astrophys. J. Suppl.}}
  \textbf{\bibinfo{volume}{148}}, \bibinfo{pages}{543--566}
  (\bibinfo{year}{2003}).

\bibitem{Lawler2001La}
\bibinfo{author}{{Lawler}, J.~E.}, \bibinfo{author}{{Bonvallet}, G.} \&
  \bibinfo{author}{{Sneden}, C.}
\newblock \bibinfo{title}{{Experimental Radiative Lifetimes, Branching
  Fractions, and Oscillator Strengths for La II and a New Determination of the
  Solar Lanthanum Abundance}}.
\newblock \emph{\bibinfo{journal}{Astrophys. J.}}
  \textbf{\bibinfo{volume}{556}}, \bibinfo{pages}{452--460}
  (\bibinfo{year}{2001}).

\bibitem{Lawler2001Eu}
\bibinfo{author}{{Lawler}, J.~E.}, \bibinfo{author}{{Wickliffe}, M.~E.},
  \bibinfo{author}{{den Hartog}, E.~A.} \& \bibinfo{author}{{Sneden}, C.}
\newblock \bibinfo{title}{{Improved Laboratory Transition Parameters forEu II
  and Application to the Solar Europium Elemental and Isotopic Composition}}.
\newblock \emph{\bibinfo{journal}{Astrophys. J.}}
  \textbf{\bibinfo{volume}{563}}, \bibinfo{pages}{1075--1088}
  (\bibinfo{year}{2001}).

\bibitem{Lawler2001Tb}
\bibinfo{author}{{Lawler}, J.~E.}, \bibinfo{author}{{Wickliffe}, M.~E.},
  \bibinfo{author}{{Cowley}, C.~R.} \& \bibinfo{author}{{Sneden}, C.}
\newblock \bibinfo{title}{{Atomic Transition Probabilities in Tb II with
  Applications to Solar and Stellar Spectra}}.
\newblock \emph{\bibinfo{journal}{Astrophys. J. Suppl.}}
  \textbf{\bibinfo{volume}{137}}, \bibinfo{pages}{341--349}
  (\bibinfo{year}{2001}).

\bibitem{Lawler2006Sm}
\bibinfo{author}{{Lawler}, J.~E.}, \bibinfo{author}{{Den Hartog}, E.~A.},
  \bibinfo{author}{{Sneden}, C.} \& \bibinfo{author}{{Cowan}, J.~J.}
\newblock \bibinfo{title}{{Improved Laboratory Transition Probabilities for Sm
  II and Application to the Samarium Abundances of the Sun and Three
  r-Process-rich, Metal-poor Stars}}.
\newblock \emph{\bibinfo{journal}{Astrophys. J. Suppl.}}
  \textbf{\bibinfo{volume}{162}}, \bibinfo{pages}{227--260}
  (\bibinfo{year}{2006}).

\bibitem{Sneden2009}
\bibinfo{author}{{Sneden}, C.}, \bibinfo{author}{{Lawler}, J.~E.},
  \bibinfo{author}{{Cowan}, J.~J.}, \bibinfo{author}{{Ivans}, I.~I.} \&
  \bibinfo{author}{{Den Hartog}, E.~A.}
\newblock \bibinfo{title}{{New Rare Earth Element Abundance Distributions for
  the Sun and Five r-Process-Rich Very Metal-Poor Stars}}.
\newblock \emph{\bibinfo{journal}{Astrophys. J. Suppl.}}
  \textbf{\bibinfo{volume}{182}}, \bibinfo{pages}{80--96}
  (\bibinfo{year}{2009}).

\bibitem{McCully2017}
\bibinfo{author}{{McCully}, C.} \emph{et~al.}
\newblock \bibinfo{title}{{The Rapid Reddening and Featureless Optical Spectra
  of the Optical Counterpart of GW170817, AT 2017gfo, during the First Four
  Days}}.
\newblock \emph{\bibinfo{journal}{Astrophys. J. Lett.}}
  \textbf{\bibinfo{volume}{848}}, \bibinfo{pages}{L32} (\bibinfo{year}{2017}).

\bibitem{Chornock2017}
\bibinfo{author}{{Chornock}, R.} \emph{et~al.}
\newblock \bibinfo{title}{{The Electromagnetic Counterpart of the Binary
  Neutron Star Merger LIGO/Virgo GW170817. IV. Detection of Near-infrared
  Signatures of r-process Nucleosynthesis with Gemini-South}}.
\newblock \emph{\bibinfo{journal}{Astrophys. J. Lett.}}
  \textbf{\bibinfo{volume}{848}}, \bibinfo{pages}{L19} (\bibinfo{year}{2017}).

\bibitem{2010IAUS..265...46S}
\bibinfo{author}{{Sneden}, C.}, \bibinfo{author}{{Cowan}, J.~J.} \&
  \bibinfo{author}{{Gallino}, R.}
\newblock \bibinfo{title}{{Constraints on the Nature of the s- and
  r-processes}}.
\newblock In \bibinfo{editor}{{Cunha}, K.}, \bibinfo{editor}{{Spite}, M.} \&
  \bibinfo{editor}{{Barbuy}, B.} (eds.) \emph{\bibinfo{booktitle}{Chemical
  Abundances in the Universe: Connecting First Stars to Planets}}, vol.
  \bibinfo{volume}{265} of \emph{\bibinfo{series}{IAU Symposium}},
  \bibinfo{pages}{46--53} (\bibinfo{year}{2010}).

\bibitem{Kurucz_gfall}
\bibinfo{title}{{Kurucz} line list}.
\newblock
  \bibinfo{howpublished}{\url{http://kurucz.harvard.edu/linelists/gfnew/gfall08oct17.dat}}.

\bibitem{Tanaka2018}
\bibinfo{author}{{Tanaka}, M.} \emph{et~al.}
\newblock \bibinfo{title}{{Properties of Kilonovae from Dynamical and
  Post-merger Ejecta of Neutron Star Mergers}}.
\newblock \emph{\bibinfo{journal}{Astron. Astrophys.}}
  \textbf{\bibinfo{volume}{852}}, \bibinfo{pages}{109} (\bibinfo{year}{2018}).

\bibitem{Karp1977}
\bibinfo{author}{Karp, A.~H.}, \bibinfo{author}{Lasher, G.},
  \bibinfo{author}{Chan, K.~L.} \& \bibinfo{author}{Salpeter, E.~E.}
\newblock \bibinfo{title}{{The opacity of expanding media - The effect of
  spectral lines}}.
\newblock \emph{\bibinfo{journal}{Astrophys. J.}}
  \textbf{\bibinfo{volume}{214}}, \bibinfo{pages}{161} (\bibinfo{year}{1977}).

\bibitem{Barnes2013b}
\bibinfo{author}{Barnes, J.} \& \bibinfo{author}{Kasen, D.}
\newblock \bibinfo{title}{{Effect of a high opacity on the light curves of
  radioactively powered transients from compact object mergers}}.
\newblock \emph{\bibinfo{journal}{Astrophys. J.}}
  \textbf{\bibinfo{volume}{775}}, \bibinfo{pages}{18} (\bibinfo{year}{2013}).

\bibitem{2017arXiv171005432S}
\bibinfo{author}{{Shappee}, B.~J.} \emph{et~al.}
\newblock \bibinfo{title}{{Early Spectra of the Gravitational Wave Source
  GW170817: Evolution of a Neutron Star Merger}}.
\newblock \emph{\bibinfo{journal}{ArXiv e-prints}}  (\bibinfo{year}{2017}).

\bibitem{2017arXiv171109638W}
\bibinfo{author}{{Waxman}, E.}, \bibinfo{author}{{Ofek}, E.},
  \bibinfo{author}{{Kushnir}, D.} \& \bibinfo{author}{{Gal-Yam}, A.}
\newblock \bibinfo{title}{{Constraints on the ejecta of the GW170817
  neutron-star merger from its electromagnetic emission}}.
\newblock \emph{\bibinfo{journal}{ArXiv e-prints}}  (\bibinfo{year}{2017}).

\bibitem{2000ApJ...530..757P}
\bibinfo{author}{{Pinto}, P.~A.} \& \bibinfo{author}{{Eastman}, R.~G.}
\newblock \bibinfo{title}{{The Physics of Type IA Supernova Light Curves. II.
  Opacity and Diffusion}}.
\newblock \emph{\bibinfo{journal}{Astrophys. J.}}
  \textbf{\bibinfo{volume}{530}}, \bibinfo{pages}{757--776}
  (\bibinfo{year}{2000}).

\bibitem{2016ascl.soft06014N}
\bibinfo{author}{{Newville}, M.} \emph{et~al.}
\newblock \bibinfo{title}{{Lmfit: Non-Linear Least-Square Minimization and
  Curve-Fitting for Python}}.
\newblock \bibinfo{howpublished}{Astrophysics Source Code Library}
  (\bibinfo{year}{2016}).

\bibitem{2013PASP..125..306F}
\bibinfo{author}{{Foreman-Mackey}, D.}, \bibinfo{author}{{Hogg}, D.~W.},
  \bibinfo{author}{{Lang}, D.} \& \bibinfo{author}{{Goodman}, J.}
\newblock \bibinfo{title}{{emcee: The MCMC Hammer}}.
\newblock \emph{\bibinfo{journal}{Proc. Ast. Soc. Pacific}} \textbf{\bibinfo{volume}{125}},
  \bibinfo{pages}{306} (\bibinfo{year}{2013}).

\bibitem{2017PASJ...69..102T}
\bibinfo{author}{{Tanaka}, M.} \emph{et~al.}
\newblock \bibinfo{title}{{Kilonova from post-merger ejecta as an optical and
  near-Infrared counterpart of GW170817}}.
\newblock \emph{\bibinfo{journal}{Proc. Ast. Soc. Japan}}
  \textbf{\bibinfo{volume}{69}}, \bibinfo{pages}{102} (\bibinfo{year}{2017}).

\bibitem{2014MNRAS.443.3134P}
\bibinfo{author}{{Perego}, A.} \emph{et~al.}
\newblock \bibinfo{title}{{Neutrino-driven winds from neutron star merger
  remnants}}.
\newblock \emph{\bibinfo{journal}{Mon. Not. R. Astron. Soc.}}
  \textbf{\bibinfo{volume}{443}}, \bibinfo{pages}{3134--3156}
  (\bibinfo{year}{2014}).

\bibitem{Sneden2003}
\bibinfo{author}{{Sneden}, C.} \emph{et~al.}
\newblock \bibinfo{title}{{The Extremely Metal-poor, Neutron Capture-rich Star
  CS 22892-052: A Comprehensive Abundance Analysis}}.
\newblock \emph{\bibinfo{journal}{Astrophys. J.}}
  \textbf{\bibinfo{volume}{591}}, \bibinfo{pages}{936--953}
  (\bibinfo{year}{2003}).

\end{thebibliography}
\end{document}